\documentclass{article}

\usepackage[eandd,final]{neurips_2026}
\usepackage{amsmath}

\usepackage{listings}
\usepackage{xcolor}

\usepackage{multirow}
\usepackage{colortbl}
\usepackage{array}
\usepackage{adjustbox}
\usepackage{natbib}
\usepackage{enumitem}
\usepackage{tcolorbox}
\tcbuselibrary{breakable}
\usepackage{ragged2e}
\usepackage{tabularx}
\newcolumntype{Y}{>{\RaggedRight\arraybackslash}X}

\definecolor{goldrow}{RGB}{255,248,220}
\definecolor{pooledrow}{RGB}{230,240,255}
\definecolor{highlight}{RGB}{255,235,205} 

\newcommand{\gold}[1]{\cellcolor{goldrow}#1}
\newcommand{\pooled}[1]{\cellcolor{pooledrow}#1}
\newcommand{\hgold}[1]{\cellcolor{highlight!70}#1}  
\newcommand{\hpooled}[1]{\cellcolor{pooledrow!155}#1}

\usepackage[utf8]{inputenc} 
\usepackage[T1]{fontenc}    
\usepackage{hyperref}       
\usepackage{url}            
\usepackage{booktabs}       
\usepackage{amsfonts}       
\usepackage{nicefrac}       
\usepackage{microtype}      
\usepackage{xcolor}         

\title{OBLIQ-Bench: Exposing \underline{O}verlooked \underline{B}ottlenecks in Modern Retrievers with \underline{L}atent and \underline{I}mplicit \underline{Q}ueries}

\author{%
  Diane Tchuindjo \quad
  Devavrat Shah \quad
  Omar Khattab \\[0.5em]
  Massachusetts Institute of Technology, Cambridge, MA \\
  \texttt{\{dianetc, devavrat, okhattab\}@mit.edu}
}

\begin{document}

\maketitle
\begin{abstract}


Retrieval benchmarks are increasingly saturating, but we argue that efficient search is far from a solved problem. We identify a class of queries we call \emph{oblique}, which seek documents that \textit{instantiate a latent pattern}, like finding all tweets that express an implicit stance, chat logs that demonstrate a particular failure mode, or transcripts that match an abstract scenario. We study three mechanisms through which obliqueness may arise 
and introduce OBLIQ-Bench, a suite of five oblique search problems over real long-tail corpora. OBLIQ-Bench exposes an overlooked asymmetry between retrieval and verification, where reasoning LLMs reliably recognize latent relevance whenever relevant documents are surfaced, but even sophisticated retrieval pipelines fail to surface most relevant documents in the first place. We hope that OBLIQ-Bench will drive research into retrieval architectures that efficiently capture latent patterns and implicit signals in large corpora.\footnote{Data: \url{https://huggingface.co/datasets/dianetc/OBLIQ-Bench}. Code, prompts, and system
rankings: \url{https://github.com/dianetc/OBLIQ-Bench}.}

\end{abstract}

\begin{figure*}[h]
\centering
\includegraphics[width=0.94\textwidth]{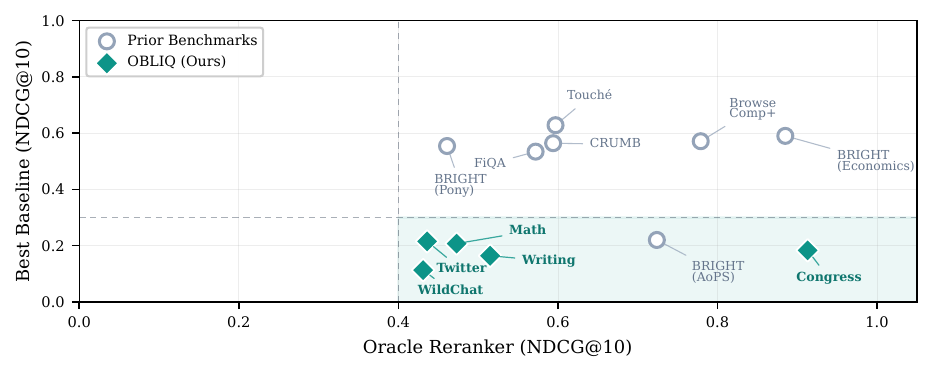}

\caption{\textbf{Compared with prior benchmarks, relevant documents in OBLIQ-Bench are easy to verify but much harder to retrieve.}
Each point on this plot is a retrieval benchmark. The $y$ axis shows the best NDCG@10 obtained by a suite of state-of-the-art retrieval systems and agentic multi-hop search pipelines. The $x$ axis shows the NDCG@10 obtained when a reasoning model re-ranks a very large pool of hard distractors infused with the gold results. Prior benchmarks mostly sit near the diagonal: strong retrievers recover much of what the reasoning model can verify. In contrast, oblique tasks fall in the lower-right: the reasoning model can identify relevant documents, but even LLM-driven retrieval systems fail to surface them. More details in Sections~\ref{sec:definition} and~\ref{sec:eval}.}


\label{fig:gap-comparison}
\end{figure*}
\section{Introduction}
\label{sec:intro}


Modern information retrieval (IR) benchmarks increasingly portray the existing IR paradigms as highly effective and hard to empirically distinguish. Across standard passage-ranking datasets like MS MARCO~\citep{msmarco}, BEIR~\citep{beir}, MTEB~\citep{mteb}, and even BRIGHT~\citep{bright}, signs of saturation are common and it is increasingly hard to observe more than a marginal difference in metrics like NDCG@$k$ across strong dense retrievers~\citep{dpr,rocketqa}, late interaction models~\citep{colbert}, and LLM-based multi-hop agents~\citep{react,karl}.


\begin{figure}[!b]
\vspace{-2mm}
\centering
\includegraphics[width=0.92\textwidth]{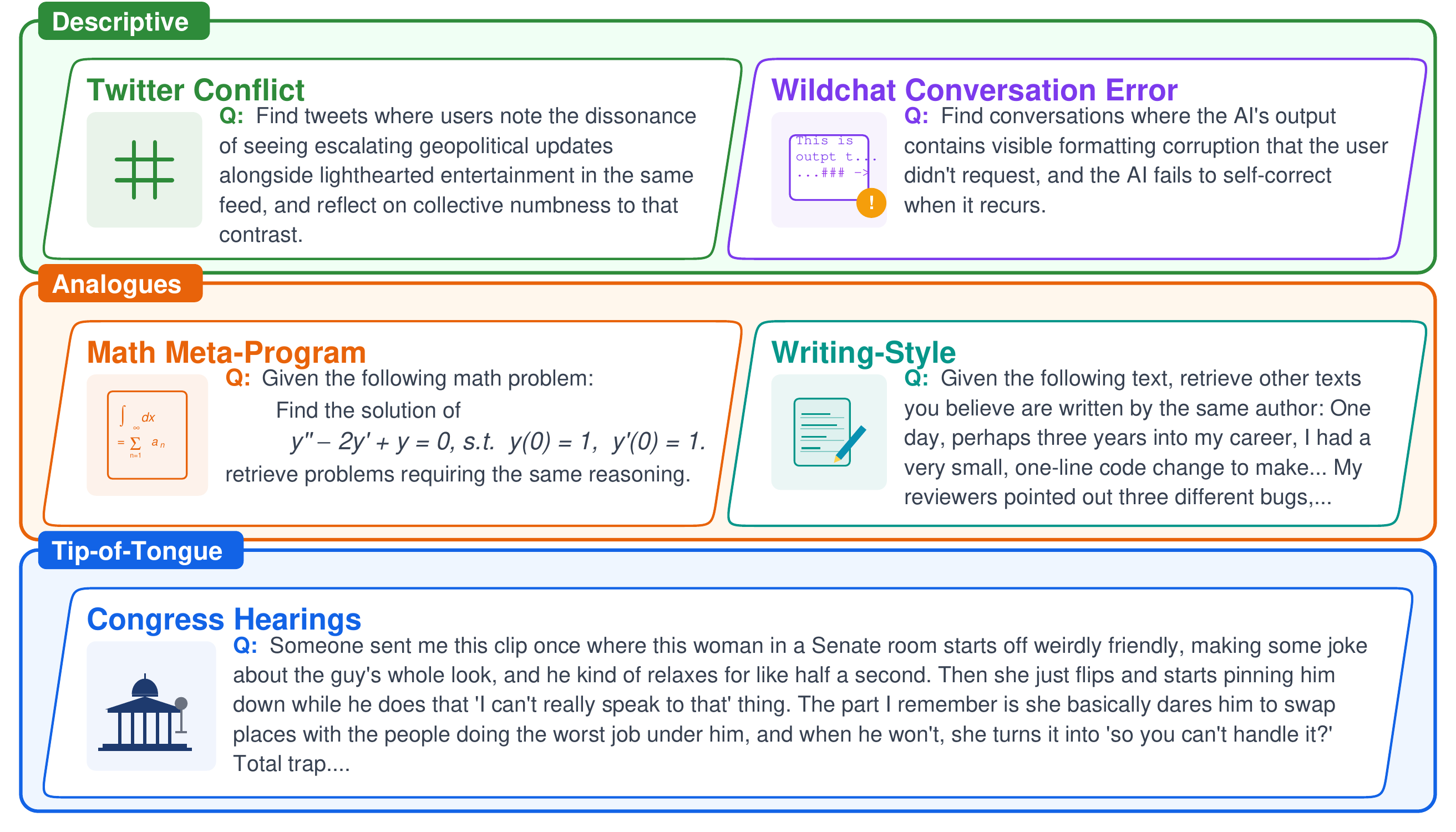}
\vspace{-3mm}
\caption{\textbf{Our five OBLIQ-Bench tasks span three types of oblique search queries.} Descriptive queries seek a latent property that can be inferred from document content, like tweets that subtly imply a detailed stance and Human--AI conversations that exhibit an implicit failure mode. Analogue queries seek all documents that share an archetype with the content of the query, despite differing in surface topic, like math problems that yield to the same proof technique or text snippets that share an author's stylistic fingerprint across topics.
Lastly, tip-of-the-tongue queries match a fuzzy recollection to a specific obscure passage, like a transcript of a Congress hearing.
In every task, relevance is verifiable by a reasoning model but hard for current retrieval systems.}
\label{fig:overview}
\vspace{-3mm}
\end{figure}

In fact, in numerous existing IR benchmarks (Figure~\ref{fig:gap-comparison}; top right), documents whose relevance can be verified by a reasoning model are, to a large extent, already easy to retrieve by efficient search algorithms.
If scalable retrievers can nearly match the ceiling set by a prohibitive frontier LLM that can read the query and document together, what is left to solve? 

We argue that this apparent saturation is an artifact of existing benchmarks, not a property of the problem of retrieval. There exist natural queries, ones that real users might pose and that ambitious downstream systems should support, on which current retrieval architectures score extremely poorly while a reasoning LLM scores substantially higher (Figure~\ref{fig:gap-comparison}; bottom right). These queries share an informal property we call \emph{obliqueness}: the attributes that determine relevance are latent  and have little or no surface expression in the document. 

Consider an analyst searching a corpus of tweets for any post $p$ that \textit{mock how users turn distant armed clashes into entertainment} without $p$ expressing that explicitly. The relevant tweets may communicate their stance through irony or the choice of what to emphasize, but will not generally overlap directly with the query. Or consider an auditor searching Human--AI conversations for cases where a model \textit{receives a formatting constraint, violates it several turns later, and never corrects itself}. The failure is evident in the transcript but might not be a topic that is  acknowledged during the conversation. Or perhaps an ML expert who is trying to decontaminate training data may want to find \textit{all problems that use the same proof strategy as a given question} despite different details. 

Toward pushing the field to tackle such ambitious retrieval challenges, we make three contributions. First, we identify oblique retrieval as an overlooked regime for evaluating search systems. We formalize the gap between retrieval failure and verification success, or the \emph{retrieval--verification asymmetry} (\S\ref{sec:definition}), as an empirical test for obliqueness, and use it to show that existing benchmarks often fail the test: they are not oblique with respect to existing methods (Figure~\ref{fig:gap-comparison}).

Second, we develop OBLIQ-Bench (\S\ref{sec:datasets}) by introducing five difficult retrieval tasks on top of five long-tail corpora (Figure~\ref{fig:overview}). To obtain high-recall relevance judgments over a large corpus without exhaustively judging every query--document pair, we propose a pipeline for starting from a human-defined lens for reading a corpus---for example, implicit stance in tweets or failure modes in conversations---and allowing a reasoning LLM to (1) annotate the entire corpus over an affordable single pass, (2) cluster documents using these extracted latent attribute annotations, and (3) generate well-scoped queries that describe each cluster. 
Figure~\ref{fig:overview} shows representative example queries that we can produce, along with their relevance judgments, in this manner. 

Finally, we evaluate lexical, dense, late interaction, and agentic retrievers across the five tasks (\S\ref{sec:eval}), and derive a set of five lessons about the existing IR landscape. All systems score poorly, and often close to zero NDCG@10, on every task. In contrast, a reasoning model applying tournament reranking over a very large pool of hard distractors (Figure~\ref{fig:scaling}) scales dramatically better with the pool size and reliably separates the relevant documents, confirming that the relevance signal exists but can be inaccessible to current approaches that do not apply joint reasoning over query and document.We release all corpora, queries, and relevance judgments, together with our construction and evaluation code, prompts, and the rankings of every system we evaluate, and hope OBLIQ-Bench drives
work on new scalable retrieval approaches that can begin to attack such oblique queries.

\section{Background and Related Work}
\label{sec:background}

We are interested in scalable retrieval over a massive corpus $\mathcal C=\{d_1,\ldots,d_N\}$. At query time, a retriever receives a query $q$ and an integer $k\ll N$, and must return an ordered list $d_{i_1},\ldots,d_{i_k}$. The retriever may spend considerable work offline, say, to precompute representations or build an index, but its per-query cost should be sublinear in the corpus size, for instance $O(N^{0.5})$ total operations with at most $O(1)$ of them being calls to large LLMs, or at least it should have such small constants that querying a large $\mathcal C$ is possible. Modern retrieval paradigms fit this setting, not only lexical, dense retrievers, and late interaction methods but also highly sophisticated LLM search agents or query rewriting pipelines.
We evaluate quality using standard IR metrics like NDCG@$k$ and Recall@$k$.

OBLIQ-Bench focuses on queries where relevance is determined by a latent relation. \textit{Oblique descriptive queries} seek documents that express a latent property implicitly, such as an implicit stance or behavioral failure mode. \textit{Oblique analogue queries} seek documents that share an abstract structure with the query, such as a proof strategy or authorial style, despite differing in surface topic. Lastly, \textit{oblique tip-of-the-tongue queries} seek an obscure passage from a partial, lossy, and abstract recollection. These mechanisms touch several lines of prior work, but in OBLIQ-Bench \textit{the bottleneck appears very early in the search process}.


For instance, existing description-based retrieval queries~\citep{descriptive} search for sentences from abstract descriptions of their content, like matching ``a change of career path'' to ``moved on to a manager career'' which is a shallower linguistic lens. ImpliRet~\citep{taghavi2025impliret} likewise moves the reasoning to the document side: its queries are simple, but relevance depends on facts stated implicitly in synthetic documents, such as a relative date or a landmark that implies a country. In OBLIQ-Bench, the latent attribute is instead an interpretive property of real documents, such as a stance, a failure mode, or a proof strategy. Existing tip-of-the-tongue and inferential queries typically seek a \textit{popular entity} like a movie, book, or celebrity~\citep{tot,lin2023decomposing,mozafari2026inferential}, like matching ``Spanish-speaking footballer holding the most titles'' to passages about Lionel Messi, which is increasingly vulnerable to query re-writing via LLM parametric knowledge. Lastly, reasoning-intensive retrieval such as BRIGHT seek evidence documents that can be hard to verify but which are not systematically harder to retrieve into a top-$k$ pool. OBLIQ-Bench seeks to make the latent relevance itself a first-stage search bottleneck and thus places the difficulty earlier (at scalably surfacing the documents in the first place),
motivating the retrieval--verification diagnostic below.
Appendix~\ref{extended-works} contains an extended related work section.

\section{Obliqueness and Retrieval--Verification Asymmetry}
\label{sec:definition}

We seek to present more ambitious challenges to retrievers than current popular benchmarks do. To do so, we must guard against tasks that appear hard only because queries are ill-posed or where it is not possible to exceed a low score. In other words, we must find tasks $t$ such that, for a set $\mathcal R$ of scalable retrieval systems of interest, the \textit{retrieval--verification gap} is large. We define this gap as follows. For quality metric $Q$, let ${\mathrm gap}(t) = V_t - R_t$ where 
$V_t = \max_{m \in \mathcal M} Q(m,t)$ represents the best quality achieved over any model $m$ in a powerful model class ${\mathcal M}$, which could ordinarily be prohibitively expensive, and  where $R_t = \max_{r \in \mathcal R} Q(r,t)$ is the quality of the best-scoring $r \in \mathcal R$, an efficient retrieval system class.

Although it can be affordable (\S\ref{sec:datasets}) to run a powerful model $m$ on every document in $\mathcal C$ once or twice, we cannot exhaustively repeat this for every query $q$. Instead, we can ask some efficient system $r \in \mathcal R$ to fetch a set $K \gg k$ of documents and then rank these unordered $K$ documents using a powerful model $m \in \mathcal M$.\footnote{This is inspired by pooling~\citep{voorhees2007trec}, TREC's standard approach for evaluating retrievers, in which ``the judge reviews only the documents [in] the union of the set of X top-retrieved documents for that topic by each [retrieval system]''.} Doing so establishes a \textit{lower bound} on the quality of the best possible system $Q(\star,t)$. With modest assumptions about the stability of $m$, we can further approximate an \textit{upper bound} on the quality of this same two-stage $\langle r, m \rangle$ system by ensuring that every missing ``gold'' document for a query $q$ is injected with the (shuffled) $K$ documents from $r$.

\begin{figure*}[t]
\centering

\begin{minipage}[t]{0.32\textwidth}
    \centering
    \includegraphics[width=\linewidth]{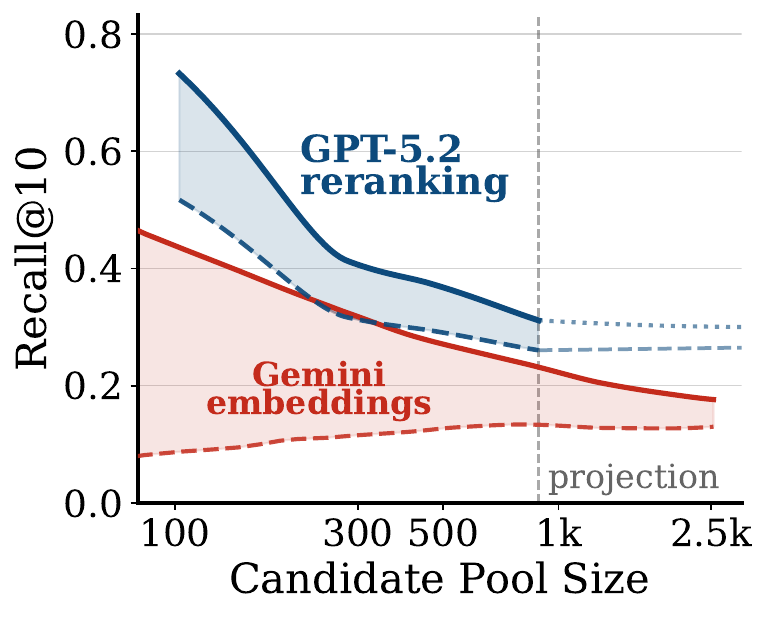}
    \centerline{\small (a) Twitter Conflict}
\end{minipage}
\hfill
\begin{minipage}[t]{0.32\textwidth}
    \centering
    \includegraphics[width=\linewidth]{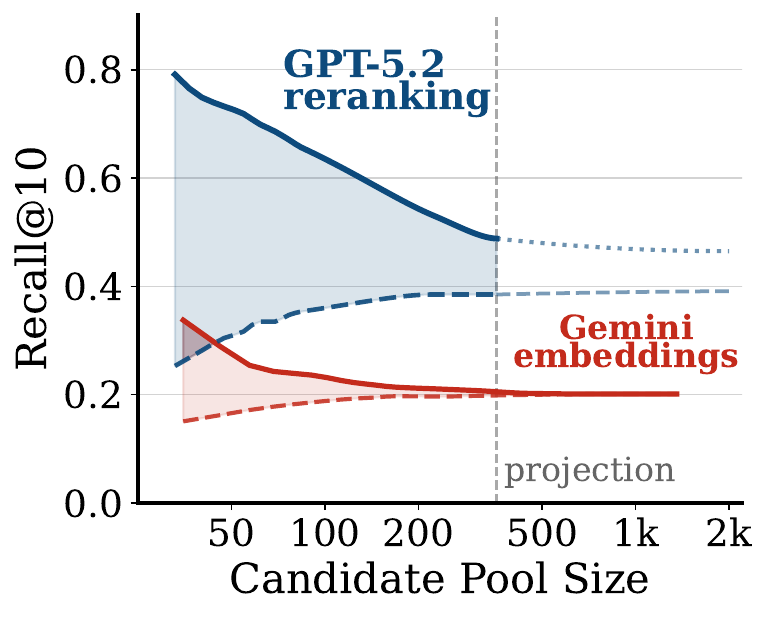}
    \centerline{\small (b) Math Meta Program}
\end{minipage}
\hfill
\begin{minipage}[t]{0.32\textwidth}
    \centering
    \includegraphics[width=\linewidth]{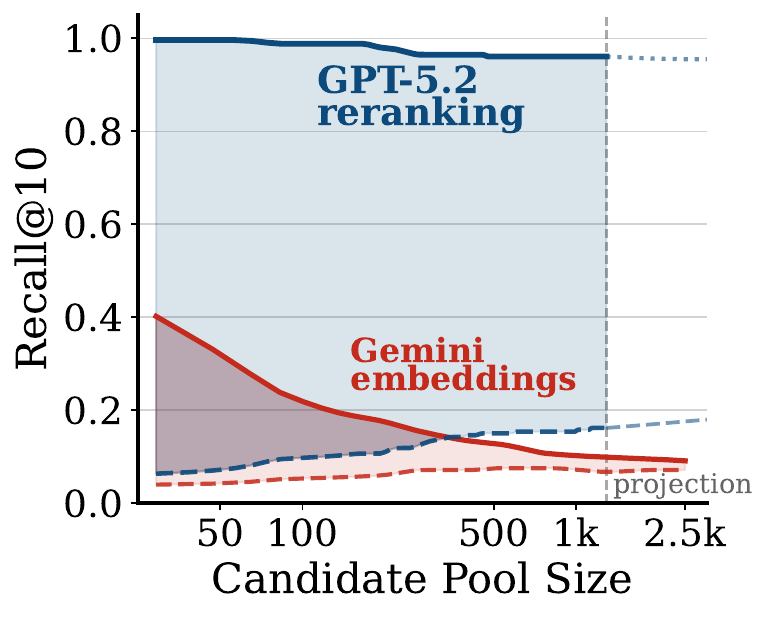}
    \centerline{\small (c) Congress Hearings}
\end{minipage}

\caption{\textbf{The retrieval--verification gap persists as hard candidate pools grow.}
Recall@10 for the state-of-the-art dense retriever Gemini-Embedding-2 and the GPT-5.2 reranker as the size $K$ of the candidate pool increases. For each model, dashed curves rank the retrieved pool as-is, giving a lower estimate that depends on the recall of the underlying pool. Solid curves inject missing gold documents into the pool before ranking, giving an upper estimate of how well the model recognizes positives once they are present. Faint and dotted GPT-5.2 segments after the vertical line are projected beyond the largest pool reranked. Across all three tasks, GPT-5.2 reranking remains far above Gemini-Embedding-2 similarity in the upper and lower settings respectively.}
\label{fig:scaling}

\vspace{-2mm}
\end{figure*}


Using these two bounds, we can begin to estimate the retrieval--verification gap of a task $t$ against a set $\mathcal R$ of state-of-the-art scalable retrieval systems. To make this estimation more reliable, we can monitor how the two bounds move as we scale $K$ and observe how this movement for a powerful model $m \in \mathcal M$ differs from how retrievers $r \in \mathcal R$ move on the same quality metric $Q$.

Figure~\ref{fig:scaling} demonstrates these patterns for a strong reasoning model $m$, which is set to GPT-5.2, and for the strongest frontier dense retriever in our experiments, Gemini-Embedding-2. The details of our listwise GPT-5.2 Tournament reranker and our other methods are discussed in \S\ref{sec:eval}. Here, we run a pool scaling experiment on three representative OBLIQ-Bench tasks: Twitter conflict (a descriptive task), Math Meta Program (an analogue task), and Congress Hearings (a tip-of-the-tongue task). For each task and each of the two methods, we evaluate Recall@10 as the candidate-pool budget increases, for two settings: standard and gold-infused.\footnote{GPT-5.2 is evaluated over pools constructed from the union of the BM25, Qwen3-Embedding-0.6B, Qwen3-Embedding-4B, and Gemini-Embedding-2 top-$k'$ outputs. The Gemini embedding model is tested
on equal pool sizes but without its own top-$k'$, because that would otherwise present it with its own hardest distractors and lower its score. Leaving out Gemini's own top results therefore favors Gemini, so the gap we report is conservative. Only the Congress task has a large distance between the upper and lower lines for GPT-5.2: since there's only a single relevant document per query in
this task, the bottom line can never exceed the (low) success rate of the pool.} Across all three tasks and in each of our two settings, GPT-5.2 reranking remains far above Gemini-Embedding-2 as $K$ grows.

Having validated this pooling approach, we apply it across many popular existing IR benchmarks and to our own five tasks in Figure~\ref{fig:gap-comparison}. For each benchmark $t$, our estimate of the verification score $V_t$ is on the x-axis and the best scalable retrieval score $R_t$ is on the y-axis. The results suggest that many existing benchmarks are not oblique. On Touch\'{e}-2020~\citep{bondarenko2020overview}, FiQA-2018~\citep{maia201818}, CRUMB (StackExchange QA; \citealt{crumb}), and BrowseComp-Plus~\citep{browsecomp-plus}, strong dense or agentic systems recover much of the score achieved by the gold-injected reranker. 
Only the AoPS slice of BRIGHT comes close to OBLIQ-Bench in our comparison which fits since AoPS relevance is itself an analogue relation. Its queries are competition problems, and its positives are solved problems that use the same theorem or technique, drawn from AoPS Wiki articles that link each technique to problems it solves~\citep{bright}. Math Meta-Program targets the same
mechanism but differs in two ways. Its documents are problem statements without solutions, so the shared reasoning must be inferred rather than read from a worked solution. And its clusters group problems by an abstract proof move chosen to span several fields, rather than by a named technique, a grain our fingerprinting prompt treats as too coarse (Appendix~\ref{sec:appendix-math}). The one
prior benchmark in Figure~1 with a comparable gap thus relies on the analogue mechanism; OBLIQ-Bench also builds tasks for the descriptive and tip-of-the-tongue mechanisms.

\section{OBLIQ-Bench}
\label{sec:datasets}

\begin{figure}[t]
\centering
 \includegraphics[width=\linewidth]{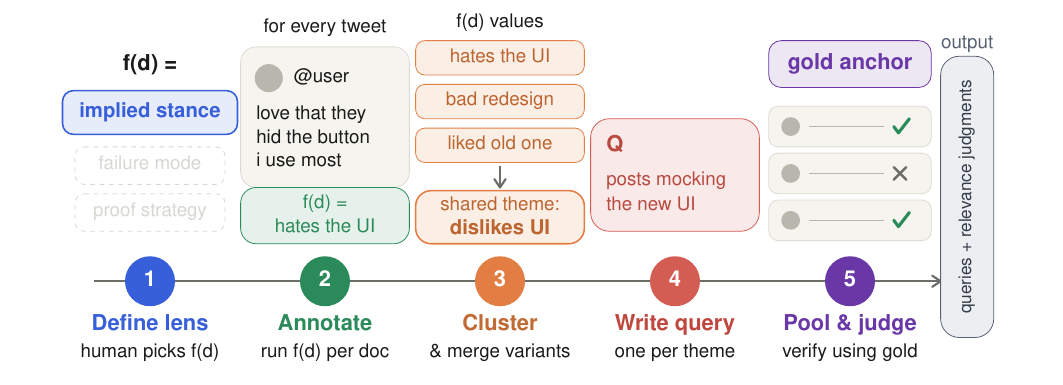}
\caption{\textbf{Construction pipeline across OBLIQ-Bench.}
A human defines a latent attribute (Stage~1). An LLM annotates documents through that lens (Stage~2), clusters attribute values (Stage~3), and generates abstract queries while forbidding source vocabulary (Stage~4). A pooling step optionally expands relevance judgments after evaluation (Stage~5). Writing-Style skips annotation and clustering because authorship is ground truth and Congress skips clustering, because each query in these two tasks targets one passage. All other tasks follow this entire pipline.}
\label{fig:pipeline}

\vspace{-2mm}
\end{figure}


We now introduce OBLIQ-Bench, a suite of five retrieval tasks designed to exhibit retrieval--verification asymmetry (\S\ref{sec:definition}) on real long-tail corpora. The tasks instantiate the three mechanisms of obliqueness from Figure~\ref{fig:overview}. 
The central construction problem in OBLIQ-Bench is: \textit{how should we obtain high-recall relevance judgments for latent properties at the scale of a large corpus, without judging every query--document pair?} If we simply asked a model to write queries for randomly sampled documents, we would risk producing artificially easy paraphrase tasks, or else having to judge the whole corpus after the fact to find the other documents that satisfy the same query.

\textbf{Steps 1--2:} Our construction pipeline (Figure~\ref{fig:pipeline}) instead begins by
\textit{extracting the value of a latent attribute} from each document $d \in \mathcal C$ in a single upfront pass. Each of our five tasks defines an unobserved, and potentially sophisticated and costly to compute, document attribute $f(d)$ that determines relevance. In our five tasks, the function $f$ is specified in natural language by a human: the stance implied by a tweet, the failure mode exhibited in a conversation, the proof strategy behind a math problem, the authorship fingerprint of a snippet, or the scenario that characterizes a transcript.

\textbf{Step 3:} Having produced $F = \{ f(d) : d \in \mathcal{C} \}$, we then cluster the indexed values in $F$, producing document groups $\mathcal G=\{G_1,\ldots,G_m\}$, where each $G_j \subseteq \mathcal C$ contains documents with near-equivalent values of $f(d)$. We do this differently for each dataset, using multiple reasoning-LLM passes and, when useful, embedding models. The important point is that these models operate on the extracted attributes $f(d)$, not directly on the raw documents. This design is deliberate: it is far easier and more reliable to extract a scoped attribute from one document, and to merge similar attribute descriptions, than to search an arbitrarily large corpus with oblique queries. These passes merge near-synonymous labels in order to allow for latent queries to emerge naturally from the corpus.

\textbf{Steps 4--5:} For each cluster $G_j$, a reasoning LLM is tasked to verbalize the abstract shared attribute among all $d \in G_j$ into a textual query $q$ while avoiding discriminative vocabulary and identifying details from the source documents. The initial positives for $q_j$ are the documents in $G_j$, after any task-specific filtering and rejudging. After running a large set of retrievers, the top-$k$ documents from all systems are pooled. For each query $q$, the reasoning LLM is tasked to read $q$, the known gold documents as anchors, and each of the unjudged candidates to identify any potential missed gold documents that can be added to the \textit{pooled} setting of the respective dataset. 

This recipe is adapted to the five tasks below, with expert human oversight throughout all steps (Appendix~\ref{app:human-review}) to ensure quality and make task-level adjustments when necessary.
 In all tasks, we use real corpora, retain hard distractors, and remove metadata that would create search shortcuts. The Writing-Style task uses authorship as an external latent label, so no LLM annotation is needed, and the Congress Hearings task uses known target passages, so each query is a recollection of one passage rather than a cluster description. Dataset statistics appear in Table~\ref{tab:dataset-stats} and task-specific prompts and filtering details are presented in Appendix~\ref{sec:appendix-twitter}--\ref{sec:appendix-congress}. For Twitter-Conflict, WildChat Errors, and Math Meta-Program, all reported numbers use audited labels that make duplicate documents consistent and re-check Math cluster membership (Appendix~\ref{app:audit}). For all five tasks, example queries appear in both Figure~\ref{fig:overview} and example queries, positive documents, and negative documents appear in Table~\ref{tab:qual-examples} in Appendix~\ref{sec:taskexamples}.

\phantomsection\label{sec:twitter}
\textbf{Twitter-Conflict: Descriptive Retrieval over Implicit Stance.}
In this task, we ask systems to find \textit{tweets that indicate a given stance on a topic and do so only implicitly}. We collect 72,122 English tweets posted between March 1 and 18, 2026 via the X API full-archive search endpoint, scoped to a single geopolitical conflict that began in February 2026. GPT-5 classifies each tweet as \texttt{explicit\_sentiment}, \texttt{news\_repost}, or \texttt{implicit}. Only the 7,918 implicit tweets can receive gold labels, while explicit and news tweets act as hard distractors. GPT-5 translates each implicit tweet into a short description of its stance; these descriptions are consolidated into canonical themes that become queries, and each theme's tweets are then rejudged, keeping those graded 2 (clearly relevant) or 1 (marginally relevant) with their grade
(Appendix~\ref{sec:appendix-twitter}). We also report pooled NDCG after asking the reasoning model to judge the top-10 retrieved results from the dense models and GPT 5.2 Tournament. 

\phantomsection\label{sec:wildchat}
\textbf{WildChat Errors: Descriptive Retrieval over LLM Interaction Transcripts.} Here, we ask systems to retrieve Human--AI conversations that exhibit a named behavioral failure mode. We draw English conversations in WildChat-4.8M~\citep{wildchat}, filter to 2025 conversations, serialize each as alternating user--assistant turns, and discard metadata such as model identity and topic annotations. GPT-5.4-nano performs a corpus-wide sweep and flags 14,743 conversations (2.9\%) for errors. We cluster the raw labels in embedding space and use GPT-5 to collapse them into canonical failure types. GPT-5 then writes the queries, and a final rejudging pass keeps the query--conversation pairs
graded 2 or 1, with their grade; we use the same pooled evaluation protocol as Twitter.

\begin{table}[t]
\small
\centering
\caption{Dataset statistics. $\mathcal{Q}$ is the number of queries, $\mathcal{D}$ is corpus size, $\mathcal{D}^{+}_{G}$/$\mathcal{D}^{+}_{P}$ are average gold/pooled positives per query, and lengths are in characters. Distributions of relevance grades and positives per query are in Appendix~\ref{app:qrels-stats}. Moreover, Twitter-Conflict, WildChat Errors, and Math Meta-Program use the audited labels of
Appendix~\ref{app:audit}.}
\vspace{-1mm}
\label{tab:dataset-stats}
\begin{tabular}{lrrrrrr}
\toprule
Dataset Name & $\mathcal{Q}$ & $\mathcal{D}$ & Avg $\mathcal{D}^{+}_{G}$ & Avg $\mathcal{D}^{+}_{P}$ & Avg $|\mathcal{Q}|_{char}$ & Avg $|\mathcal{D}|_{char}$ \\
\midrule
Twitter-Conflict & 281 & 72,122 & 11.7 & 16.6 & 304 & 212 \\
WildChat Errors & 40 & 507,729 & 28.9 & 32.1 & 291 & 4,161 \\
Math Meta Program & 151 & 3,508 & 10.7 & 11.3 & 599 & 316 \\
Writing-Style & 512 & 10,389 & 9.00 & -- & 3,057 & 2,315 \\
Congress Hearings & 254 & 213,650 & 1.00 & -- & 808 & 2,545 \\
\bottomrule
\end{tabular}
\end{table}

\phantomsection\label{sec:putnam}
\textbf{Math Meta-Program: Analogues via Shared Reasoning Technique.}
Here, the query is a problem whose solution uses a latent reasoning technique, and relevant documents are other problems that use the same technique across different mathematical topics. We draw problems and solutions from Putnam and other related undergraduate math competitions, the American Mathematical Monthly, and qualifying-exam sources. GPT-5 identifies the underlying meta-program of each solution, these labels are collapsed into canonical clusters, and one abstract query is generated for each.

\phantomsection\label{sec:writing}
\textbf{Writing-Style: Analogues via Cross-Domain Authorship.}
We test whether systems can retrieve prose by the same author when topic is deliberately varied. To do so, we manually derive snippets from 64 researchers who post across unrelated subjects, scrub names and bylines, and split articles into snippets; authorship supplies the gold label. We add length-matched distractors from LWN.net, LessWrong, and Quanta Magazine to preserve register while removing easy source-contiguity cues. Each gold snippet serves as a query, and relevant documents are other snippets by the same author. Snippets from the same source post are masked at retrieval time, forcing systems to rely on cross-topic stylistic invariants rather than local context or topic. No pooled labels are used since relevance is determined by authorship.

\phantomsection\label{sec:tot-hard}
\textbf{Congress Hearings: Tip-of-the-Tongue over Transcript Scenarios.}
This task asks systems to recover a single obscure passage from a lossy recollection. We scrape GovInfo congressional hearing transcripts from the 110th through 119th Congresses, segment them into speaker-turn passages, and supplement them with ten high-profile technology hearings, including major antitrust, social media, and child-safety testimonies. Each query describes the target exchange's dynamic, emotional register, or rhetorical move while omitting names, dates, committees, and verbatim transcript phrasing. A hardening pass removes remaining identifiers. Each query has one gold passage.



\section{Evaluation \& Analyses}
\label{sec:eval}

Having produced OBLIQ-Bench and validated that its query--document relevance relations are verifiable to a powerful reasoning model that operates as a ranker, we now ask what oblique retrieval tasks can teach us about the existing methods for scalable search.

\begin{table}[t]
\centering
\small
\caption{Results on the descriptive style queries. Namely, the Twitter-Conflict and WildChat tasks. Each metric shows \colorbox{goldrow}{Gold (G)} / \colorbox{pooledrow}{Pooled (P)} evaluation. \colorbox{highlight}{Key metrics} highlighted.}
\vspace{-1mm}

\label{tab:descriptive-results}
\begin{adjustbox}{max width=0.87\textwidth}
\begin{tabular}{ll cc cc ccc}
\toprule
& & \multicolumn{2}{c}{\cellcolor{highlight}\textbf{NDCG@10}} & \multicolumn{2}{c}{\textbf{NDCG@50}} & \multicolumn{3}{c}{\textbf{Recall}} \\
\cmidrule(lr){3-4} \cmidrule(lr){5-6} \cmidrule(lr){7-9}
\textbf{Dataset} & \textbf{Pipeline} & \hgold{\textbf{G}} & \hpooled{\textbf{P}} & \gold{G} & \pooled{P} & \cellcolor{highlight}\textbf{@10} & @50 & @100 \\
\midrule
\multirow{8}{*}{\rotatebox{90}{\parbox{2.2cm}{\centering\textbf{Twitter}}}}
& BM25 & \hgold{.000} & \hpooled{.002} & \gold{.001} & \pooled{.004} & \cellcolor{highlight!30}.000/.002 & .002/.007 & .002/.011 \\
& LateOn 0.1B & \hgold{.004} & \hpooled{.010} & \gold{.008} & \pooled{.017} & \cellcolor{highlight!30}.003/.009 & .014/.026 & .025/.049 \\
& Qwen3-Embed-0.6B & \hgold{.012} & \hpooled{.016} & \gold{.021} & \pooled{.037} & \cellcolor{highlight!30}.009/.009 & .035/.061 & .057/.092 \\
& Qwen3-Embed-4B & \hgold{.037} & \hpooled{.047} & \gold{.060} & \pooled{.094} & \cellcolor{highlight!30}.038/.037 & .098/.146 & .137/.196 \\
& Gemini-Embedding-2 & \hgold{.075} & \hpooled{.135} & \gold{.107} & \pooled{.240} & \cellcolor{highlight!30}.071/.093 & .166/.364 & .229/.432 \\
\cmidrule{2-9}
& GPT-5.2 Query Rewriter & \hgold{.072} & \hpooled{.142} & \gold{.106} & \pooled{.200} & \cellcolor{highlight!30}.070/.109 & .165/.271 & .215/.359\\
& GPT-5.2 Multi-Hop Agent & \hgold{.155} & \hpooled{.226} & \gold{.175} & \pooled{.257} & \cellcolor{highlight!30}.141/.168 & .210/.289 & .223/.307 \\
\cmidrule{2-9}
& Oracle GPT-5.2 Tournament & \hgold{.345} & \hpooled{.478} & \gold{.436} & \pooled{.570} & \cellcolor{highlight!30}.304/.345 & .580/.664 & .742/.797 \\
\midrule
\multirow{8}{*}{\rotatebox{90}{\parbox{2.2cm}{\centering\textbf{WildChat}}}}
& BM25 & \hgold{.003} & \hpooled{.006} & \gold{.003} & \pooled{.006} & \cellcolor{highlight!30}.003/.005 & .003/.005 & .003/.005 \\
& LateOn 0.1B & \hgold{.006} & \hpooled{.007} & \gold{.011} & \pooled{.013} & \cellcolor{highlight!30}.005/.006 & .021/.019 & .041/.038 \\
& Qwen3-Embed-0.6B & \hgold{.056} & \hpooled{.074} & \gold{.056} & \pooled{.064} & \cellcolor{highlight!30}.044/.045 & .053/.056 & .055/.058 \\
& Qwen3-Embed-4B & \hgold{.030} & \hpooled{.060} & \gold{.033} & \pooled{.060} & \cellcolor{highlight!30}.029/.046 & .049/.071 & .072/.085 \\
& Gemini-Embedding-2 & \hgold{.057} & \hpooled{.099} & \gold{.074} & \pooled{.143} & \cellcolor{highlight!30}.053/.065 & .106/.197 & .129/.232 \\
\cmidrule{2-9}
& GPT-5.2 Query Rewriter & \hgold{.063} & \hpooled{.094} & \gold{.070} & \pooled{.107} & \cellcolor{highlight!30}.039/.042 & .088/.129 & .103/.150 \\
& GPT-5.2 Multi-Hop Agent & \hgold{.066} & \hpooled{.108} & \gold{.087} & \pooled{.121} & \cellcolor{highlight!30}.050/.069 & .088/.115 & .093/.120 \\
\cmidrule{2-9}
& Oracle GPT-5.2 Tournament & \hgold{.399} & \hpooled{.435} & \gold{.545} & \pooled{.570} & \cellcolor{highlight!30}.300/.302 & .660/.653 & .828/.831 \\
\bottomrule
\end{tabular}
\end{adjustbox}
\vspace{-2mm}
\end{table}

We evaluate seven systems spanning five architecture families. For efficient single-stage retrieval, we evaluate lexical, dense, and late interaction approaches. We test BM25~\citep{bm25} as the lexical baseline with default parameters in the \texttt{rank-bm225} Python library~\citep{rank_bm25}.For dense retrieval, we use three popular state-of-the-art embedding models that index the corpus offline and retrieve by cosine similarity: \textbf{Gemini Embedding~2}, \textbf{Qwen3-Embedding-0.6B}, and \textbf{Qwen3-Embedding-4B}~\citep{geminiembed,qwen3embedding}. We also include results for a recent multi-vector late interaction model, \textbf{LateOn}~\citep{lateon}, with 149M parameters using PyLate~\citep{pylate} for indexing with the PLAID~\citep{plaid} algorithm.

For agentic retrieval, we evaluate two types of pipelines, both of which pair GPT-5.2 reasoning with Gemini-Embedding-2 dense retrieval. First, we evaluate a single-hop \textbf{GPT-5.2 Query Rewriter}, which tasks GPT-5.2 to reformulate the query once into a retrieval-optimized form, after which Gemini-Embedding-2 retrieves the top 1,000 documents using the rewritten query. Second, we evaluate a \textbf{GPT-5.2 Multi-Hop Agent}, which iteratively applies query decomposition and document reading. In particular, in each hop for a total of four hops, GPT-5.2 generates a search query conditioned on the original query and any notes accumulated from prior hops, Gemini-Embedding-2 retrieves the top-25 documents per hop, and GPT-5.2 reads each batch, selects pertinent candidates, and extracts observations to inform the next round. After four hops, GPT-selected candidates are promoted to top positions and the remaining retrieved-but-unselected documents fill the tail.

As discussed in \S\ref{sec:definition}, we establish an approximate bound on the attainable quality for each task using a powerful reasoning model, namely GPT-5.2, as a \textit{listwise} ranker~\citep{ma2023zero,sun2023is,chen2025tourrank}. Here, we construct a candidate pool by taking the union of approximately $300$ top-$k$ results from the union of Gemini-Embedding-2 and both Qwen3-Embed models along with an injection of gold documents, per query. Refer to Figure~\ref{fig:scaling} for a detailed discussion of this process. We refer to this system as \textbf{Oracle GPT-5.2 Tournament}. The procedure is a tournament-style listwise reranking algorithm, designed to expose GPT-5.2 only to small shuffled batches while focusing more compute on producing an approximate global head ranking. Appendix~\ref{sec:tournament} presents the details of the algorithm we use for re-ranking arbitrarily large sets of documents.

\begin{table}[t]
\centering
\small
\caption{Results on the analogue queries. Namely, the Math Meta-Program and Writing-Style tasks. Math shows \colorbox{goldrow}{Gold (G)} / \colorbox{pooledrow}{Pooled (P)}; Writing uses a single gold annotation. Compared to the other tables of results, here Recall@100 covers a disproportionately large fraction of each corpus, since the two corpora have only 3.5k and 10k documents respectively. 
}
\label{tab:analogue-results}
\begin{adjustbox}{max width=0.87\textwidth}
\begin{tabular}{ll cc cc ccc}
\toprule
& & \multicolumn{2}{c}{\cellcolor{highlight}\textbf{NDCG@10}} & \multicolumn{2}{c}{\textbf{NDCG@50}} & \multicolumn{3}{c}{\textbf{Recall}} \\
\cmidrule(lr){3-4} \cmidrule(lr){5-6} \cmidrule(lr){7-9}
\textbf{Dataset} & \textbf{Pipeline} & \hgold{\textbf{G}} & \hpooled{\textbf{P}} & \gold{G} & \pooled{P} & \cellcolor{highlight}\textbf{@10} & @50 & @100 \\
\midrule
\multirow{12}{*}{\rotatebox{90}{\parbox{3cm}{\centering\textbf{Math}}}}
& BM25 & \hgold{.027} & \hpooled{.034} & \gold{.039} & \pooled{.048} & \cellcolor{highlight!30}.030/.034 & .073/.077 & .100/.105 \\
& LateOn 0.1B & \hgold{.104} & \hpooled{.119} & \gold{.144} & \pooled{.164} & \cellcolor{highlight!30}.098/.107 & .236/.256 & .307/.327 \\
& Qwen3-Embed-0.6B & \hgold{.119} & \hpooled{.146} & \gold{.161} & \pooled{.189} & \cellcolor{highlight!30}.085/.103 & .251/.274 & .347/.369 \\
& Qwen3-Embed-4B & \hgold{.098} & \hpooled{.130} & \gold{.132} & \pooled{.162} & \cellcolor{highlight!30}.089/.110 & .200/.224 & .287/.318 \\
& Gemini-Embedding-2 & \hgold{.149} & \hpooled{.195} & \gold{.189} & \pooled{.230} & \cellcolor{highlight!30}.141/.175 & .292/.327 & .403/.432 \\
\cmidrule{2-9}
& GPT-5.2 Query Rewriter & \hgold{.145} & \hpooled{.183} & \gold{.212} & \pooled{.250} & \cellcolor{highlight!30}.153/.174 & .361/.389 & .456/.482 \\
& GPT-5.2 Multi-Hop Agent & \hgold{.159} & \hpooled{.199} & \gold{.220} & \pooled{.260} & \cellcolor{highlight!30}.160/.175 & .342/.365 & .420/.444 \\
\cmidrule{2-9}
& Oracle GPT-5.2 Tournament & \hgold{.273} & \hpooled{.317} & \gold{.399} & \pooled{.440} & \cellcolor{highlight!30}.304/.313 & .635/.643 & .806/.811 \\
& Oracle GPT-5.2 Tournament+Soln & \hgold{.443} & \hpooled{.479} & \gold{.566} & \pooled{.597} & \cellcolor{highlight!30}.455/.458 & .798/.804 & .915/.916 \\
\midrule
\multirow{8}{*}{\rotatebox{90}{\parbox{2.2cm}{\centering\textbf{Writing}}}}
& BM25 & \multicolumn{2}{c}{\cellcolor{highlight!30}.077} & \multicolumn{2}{c}{.114} & \cellcolor{highlight!30}.062 & .154 & .208 \\
& LateOn 0.1B & \multicolumn{2}{c}{\cellcolor{highlight!30}.105} & \multicolumn{2}{c}{.149} & \cellcolor{highlight!30}.087 &  .197 & .263  \\
& Qwen3-Embed-0.6B & \multicolumn{2}{c}{\cellcolor{highlight!30}.046} & \multicolumn{2}{c}{.060} & \cellcolor{highlight!30}.040 & .075 & .103 \\
& Qwen3-Embed-4B & \multicolumn{2}{c}{\cellcolor{highlight!30}.033} & \multicolumn{2}{c}{.046} & \cellcolor{highlight!30}.026 & .058 & .080 \\
& Gemini-Embedding-2 & \multicolumn{2}{c}{\cellcolor{highlight!30}.164} & \multicolumn{2}{c}{.220} & \cellcolor{highlight!30}.132 & .275 & .357 \\
\cmidrule{2-9}
& GPT-5.2 Query Rewriter & \multicolumn{2}{c}{\cellcolor{highlight!30}.027} & \multicolumn{2}{c}{.036} & \cellcolor{highlight!30}.022 & .044 & .064 \\
& GPT-5.2 Multi-Hop Agent & \multicolumn{2}{c}{\cellcolor{highlight!30}.061} & \multicolumn{2}{c}{.059} & \cellcolor{highlight!30}.034 & .037 & .038 \\
\cmidrule{2-9}
& Oracle GPT-5.2 Tournament & \multicolumn{2}{c}{\cellcolor{highlight!30}.515} & \multicolumn{2}{c}{.603} & \cellcolor{highlight!30}.449 & .686 & .797 \\
\bottomrule
\vspace{-2mm}
\end{tabular}
\end{adjustbox}
\end{table}

\subsection{Results by Mechanism}
\label{sec:results-by-mechanism}

\textbf{Descriptive queries.}
Table~\ref{tab:descriptive-results} shows a large retrieval--verification gap on both descriptive tasks. On Twitter-Conflict, Gemini-Embedding-2 is the strongest single-stage retriever, reaching .135 pooled NDCG@10. The GPT-5.2 Query Rewriter is only marginally better, but the multi-hop agent improves more substantially, reaching .226 pooled NDCG@10. The oracle GPT-5.2 Tournament reaches .478 pooled NDCG@10. WildChat is similar in the size of the gap: the best non-oracle systems remain between .099 and .108 pooled NDCG@10, while the oracle tournament reaches .435. However, multi-hop search helps WildChat much less, plausibly because a tweet-level stance can sometimes be approached through a few alternative reformulations, whereas a conversation failure is an even more latent behavior that is distributed across many turns, and hence is hard to actively search for.

\textbf{Analogue queries.}
Table~\ref{tab:analogue-results} shows that our analogue retrieval tasks are hard for different reasons. On Math Meta-Program, LateOn, Qwen3-Embed-0.6B, Gemini-Embedding-2, and the two GPT-5.2 retrieval pipelines all recover some signal, with the multi-hop agent reaching .199 pooled NDCG@10. The standard GPT-5.2 Tournament reaches .317, and when the tournament is also given solutions, however, it rises to .479 pooled NDCG@10. This confirms that the relevance relation depends on the (typically latent) proof structure. In contrast, on Writing-Style, BM25 beats both Qwen embedding models, LateOn beats BM25, and Gemini is the best single-stage retriever at .164 NDCG@10. This time, GPT-5.2 rewriting and multi-hop pipelines hurt substantially, perhaps because reformulating a prose snippet into retrieval keywords tends to preserve topic while destroying the stylistic signal that is the substance of retrieval for Writing-Style. The tournament, which can read query and candidate side by side, reaches .515 NDCG@10.

\textbf{Tip-of-the-tongue queries.}
Table~\ref{tab:tot-results} gives the starkest version of the retrieval--verification asymmetry. Congress Hearings has exactly one gold passage per query, so Recall@$k$ directly measures whether the target passage has been surfaced at all. BM25 never finds it at rank 10, Gemini-Embedding-2 reaches only .079 Recall@10, and here the extremely small LateOn at 149M parameters outperforms every other single-stage retriever and reaches .102. This makes sense: being able to match local alignment between a transcript and the abstract scenario in the query could plausibly benefit from multi-vector late interaction. The GPT-5.2 Query Rewriter is essentially tied with LateOn, while the multi-hop agent improves to .185 Recall@10 and then plateaus: its Recall@10, Recall@50, and Recall@100 are all .185. In other words, the agent likely retrieves the target passage during one of its hops and places it near the top, and hence looking deeper in its final ranking does not uncover many additional successes. By contrast, the tournament reaches .913 NDCG@10 and perfect Recall@100.

\begin{table}[t]
\scriptsize
\centering
\caption{Tip-of-tongue query results on Congress Hearings. Each query has exactly one gold passage; no pooled annotations are needed. \colorbox{highlight}{Key metrics} highlighted.}
\label{tab:tot-results}
\begin{tabular}{lcccccc}
\toprule
& \multicolumn{2}{c}{\textbf{NDCG}} & \multicolumn{3}{c}{\textbf{Recall}} \\
\cmidrule(lr){2-3} \cmidrule(lr){4-6}
\textbf{Pipeline} & \cellcolor{highlight}\textbf{@10} & @50 & \cellcolor{highlight}\textbf{@10} & @50 & @100 \\
\midrule
\multicolumn{6}{l}{\textit{Congress Hearings}} \\
\midrule
BM25 & \cellcolor{highlight!30}.000 & .002 & \cellcolor{highlight!30}.000 & .008 & .016 \\
LateOn 0.1B & \cellcolor{highlight!30}.083 & .094 &  \cellcolor{highlight!30}.102 & .149 & .185  \\
Qwen3-Embed-0.6B & \cellcolor{highlight!30}.006 & .014 & \cellcolor{highlight!30}.012 & .047 & .055 \\
Qwen3-Embed-4B & \cellcolor{highlight!30}.040 & .047 & \cellcolor{highlight!30}.063 & .096 & .122 \\
Gemini-Embedding-2 & \cellcolor{highlight!30}.059 & .066 & \cellcolor{highlight!30}.079 & .114 & .126 \\
\midrule
GPT-5.2 Query Rewriter & \cellcolor{highlight!30}.084 & .092 & \cellcolor{highlight!30}.102 & .134 & .158 \\
GPT-5.2 Multi-Hop Agent & \cellcolor{highlight!30}.183 & .183 & \cellcolor{highlight!30}.185 & .185 & .185 \\
\midrule
Oracle GPT-5.2 Tournament & \cellcolor{highlight!30}.913 & .919 & \cellcolor{highlight!30}.957 & .988 & 1.00 \\
\bottomrule
\end{tabular}
\end{table}

 \subsection{Lessons}
\label{sec:lessons}

\textbf{There is substantial headroom, but even the verifier is not perfect.}
Across all five tasks, the best non-oracle retrieval pipeline still score very poorly on NDCG@10 and trail considerably behind GPT-5.2 Tournament, though the gap is smaller on the Math task. A paired bootstrap over queries shows that this gap is not driven by a few queries on any tasks (Appendix~\ref{app:bootstrap}). The oracle reranker generally does well, though it is not perfect. We hope that this motivates and enables new retrieval architectures that can break through the bottleneck of these types of oblique queries and also work that seeks to improve the quality/cost tradeoffs of the oracle ranker (Appendix~\ref{sec:tournament}) that we use.

\textbf{Dense retrievers vary widely, but all of them fall short.}
Google's Gemini-Embedding-2 is the strongest dense retriever on every task, and in several cases by a large margin. At the same time, the Qwen results are not monotonic in model size: Qwen3-Embed-0.6B beats Qwen3-Embed-4B on WildChat, Math, and Writing-Style, perhaps due to the subtle ranking criteria in oblique queries. Given the success of the proprietary Gemini-Embedding-2 model, we might infer that this is a signal for the value of the training data for such non-conventional retrieval problems.

\textbf{Lexical and late-interaction systems fail in different ways.}
BM25 is mostly not competitive, which is unsurprising given that OBLIQ-Bench was built to remove direct lexical shortcuts. The one partial exception is Writing-Style, where BM25 beats both Qwen embedding models, perhaps because authorial style leaves some subtle lexical residue even across topics. LateOn shows the opposite profile: it is quite weak on the descriptive tasks, but punches well above its size on Math, Writing-Style, and Congress Hearings. This is especially clear on Congress, where a 149M-parameter late-interaction model is the best single-stage retriever. 
On the descriptive Twitter-Conflict task, LateOn often retrieves semantically natural topical matches that state the requested stance \textit{explicitly}, but those are distractors because the model fails to interpret the instruction seeking only implicit matches. Token-level matching seems useful when the query and document share local scenario structure, but is not necessarily sufficient to make up for the lack of suitable training data when the relevant property is a diffuse stance or behavioral failure, like when relevance depends on nuanced instruction following at retrieval time.

\textbf{Agentic search helps only when obliqueness can be translated into heuristic search actions.}
The multi-hop agent helps on Twitter and Congress, helps modestly on Math, barely helps on WildChat, and hurts considerably on Writing-Style. Iterative reformulation can help when the latent target can be approached through several alternative phrasings, as in implicit tweet stances or lossy recollections of a congressional exchange. It does not appear to help when the signal is distributed across a long conversation, and it can actively damage retrieval when the signal is orthogonal to topic, as in Writing-Style. Query rewriting might hence not be a silver-bullet solution to obliqueness.

\textbf{Offline indexing recovers only part of the value of knowing the latent attribute.}
A retriever may spend LLM computation offline to describe each document before any query
arrives. On Math, we compare such a query-agnostic index, built from Opus~5.5 descriptions, with
an oracle index of the fingerprints produced during benchmark construction
(Appendix~\ref{offline-indexing}). With Qwen3-Embedding-4B, the fingerprints raise NDCG@10
from .098 to .276, an upper bound since they seeded the relevance labels, while the offline
descriptions reach .138, recovering 23--35\% of the oracle's gain across metrics.

\section{Conclusion}

OBLIQ-Bench is built around a simple idea: as reasoning models become increasingly more powerful by leveraging highly expressive architectures and scaling inference-time compute up for processing each of their prompts, we can begin to define very hard retrieval problems where reasoning LLMs have no trouble recognizing even very subtle cases of relevant documents when they are shown, but current scalable retrieval systems cannot surface them from the corpus in the first place. OBLIQ-Bench seeks to put the search bottleneck into first-stage retrieval. This can, as far as our results suggest, be resistant to existing approaches for query rewriting, agentic multi-hop search, and bigger and better embedding models. Across five long-tail tasks, we find this pattern for implicit stances, behavioral failures, proof strategies, writing styles, and lossy recollections. Our construction pipeline turns these latent properties into scalable relevance judgments and the results suggest that the next frontier for retrieval might need to be architectures that make latent document attributes available at search time. Lastly, we acknowledge the use of assistance from generative AI tools, with all outputs reviewed and edited by the authors, in the preparation of portions of this paper.

\subsection*{Acknowledgments}

This work used Expanse GPU at the San Diego Supercomputer Center (SDSC) through allocation CIS250733 from the Advanced Cyberinfrastructure Coordination Ecosystem: Services \& Support (ACCESS; \citealt{ACCESS_PEARC23}) program, which is supported by U.S. National Science Foundation grants $\#2138259$, $\#2138286$, $\#2138307$, $\#2137603$, and $\#2138296$. This work was supported in part by MIT OCP project 6954031 and \texttt{cmpnd.ai}, and used OpenAI API credits gifted by Laude Institute and Mixedbread. We thank Antoine Chaffin and Benjamin Clavié for offering feedback on a late version of this manuscript.

\newpage
\bibliographystyle{ACM-Reference-Format}
 \bibliography{custom}

@inproceedings{caplan-goldwasser-2025-conceptcarve,
    title = "{C}oncept{C}arve: Dynamic Realization of Evidence",
    author = "Caplan, Eylon  and
      Goldwasser, Dan",
    editor = "Che, Wanxiang  and
      Nabende, Joyce  and
      Shutova, Ekaterina  and
      Pilehvar, Mohammad Taher",
    booktitle = "Proceedings of the 63rd Annual Meeting of the {A}ssociation for {C}omputational {L}inguistics (Volume 1: Long Papers)",
    month = jul,
    year = "2025",
    address = "Vienna, Austria",
    publisher = "Association for Computational Linguistics",
    url = "https://aclanthology.org/2025.acl-long.1014/",
    doi = "10.18653/v1/2025.acl-long.1014",
    pages = "20792--20809",
    ISBN = "979-8-89176-251-0"
}

@inproceedings{taghavi2025impliret,
  title     = {{ImpliRet}: Benchmarking the Implicit Fact Retrieval Challenge},
  author    = {Taghavi, Zeinab Sadat and Modarressi, Ali and Ma, Yunpu and Sch{\"u}tze, Hinrich},
  booktitle = {Proceedings of the 2025 Conference on Empirical Methods in Natural Language Processing},
  publisher = {Association for Computational Linguistics},
  year      = {2025},
  doi       = {10.18653/v1/2025.emnlp-main.1685}
}

@misc{fresh,
      title={FreshStack: Building Realistic Benchmarks for Evaluating Retrieval on Technical Documents}, 
      author={Nandan Thakur and Jimmy Lin and Sam Havens and Michael Carbin and Omar Khattab and Andrew Drozdov},
      year={2025},
      eprint={2504.13128},
      archivePrefix={arXiv},
      primaryClass={cs.IR},
      url={https://arxiv.org/abs/2504.13128}, 
}

@inproceedings{legal,
author = {Zheng, Lucia and Guha, Neel and Arifov, Javokhir and Zhang, Sarah and Skreta, Michal and Manning, Christopher D. and Henderson, Peter and Ho, Daniel E.},
title = {A Reasoning-Focused Legal Retrieval Benchmark},
year = {2025},
isbn = {9798400714214},
publisher = {Association for Computing Machinery},
address = {New York, NY, USA},
url = {https://doi.org/10.1145/3709025.3712219},
doi = {10.1145/3709025.3712219},
booktitle = {Proceedings of the 2025 Symposium on Computer Science and Law},
pages = {169–193},
numpages = {25},
location = {Munich, Germany},
series = {CSLAW '25}
}

@inproceedings{tot,
author = {Arguello, Jaime and Ferguson, Adam and Fine, Emery and Mitra, Bhaskar and Zamani, Hamed and Diaz, Fernando},
title = {Tip of the Tongue Known-Item Retrieval: A Case Study in Movie Identification},
year = {2021},
isbn = {9781450380553},
publisher = {Association for Computing Machinery},
address = {New York, NY, USA},
url = {https://doi.org/10.1145/3406522.3446021},
doi = {10.1145/3406522.3446021},
booktitle = {Proceedings of the 2021 Conference on Human Information Interaction and Retrieval},
pages = {5–14},
numpages = {10},
location = {Canberra ACT, Australia},
series = {CHIIR '21}
}

@inproceedings{react,
  title = {{ReAct}: Synergizing Reasoning and Acting in Language Models},
  author = {Yao, Shunyu and Zhao, Jeffrey and Yu, Dian and Du, Nan and Shafran, Izhak and Narasimhan, Karthik and Cao, Yuan},
  booktitle = {International Conference on Learning Representations (ICLR) },
  year = {2023},
  html = {https://arxiv.org/abs/2210.03629},
}

@inproceedings{rocketqa,
    title = "{R}ocket{QA}: An Optimized Training Approach to Dense Passage Retrieval for Open-Domain Question Answering",
    author = "Qu, Yingqi  and
      Ding, Yuchen  and
      Liu, Jing  and
      Liu, Kai  and
      Ren, Ruiyang  and
      Zhao, Wayne Xin  and
      Dong, Daxiang  and
      Wu, Hua  and
      Wang, Haifeng",
    editor = "Toutanova, Kristina  and
      Rumshisky, Anna  and
      Zettlemoyer, Luke  and
      Hakkani-Tur, Dilek  and
      Beltagy, Iz  and
      Bethard, Steven  and
      Cotterell, Ryan  and
      Chakraborty, Tanmoy  and
      Zhou, Yichao",
    booktitle = "Proceedings of the 2021 Conference of the North American Chapter of the Association for Computational Linguistics: Human Language Technologies",
    month = jun,
    year = "2021",
    address = "Online",
    publisher = "Association for Computational Linguistics",
    url = "https://aclanthology.org/2021.naacl-main.466/",
    doi = "10.18653/v1/2021.naacl-main.466",
    pages = "5835--5847"
}

@inproceedings{mteb,
    title = "{MTEB}: Massive Text Embedding Benchmark",
    author = "Muennighoff, Niklas  and
      Tazi, Nouamane  and
      Magne, Loic  and
      Reimers, Nils",
    editor = "Vlachos, Andreas  and
      Augenstein, Isabelle",
    booktitle = "Proceedings of the 17th Conference of the European Chapter of the Association for Computational Linguistics",
    month = may,
    year = "2023",
    address = "Dubrovnik, Croatia",
    publisher = "Association for Computational Linguistics",
    url = "https://aclanthology.org/2023.eacl-main.148/",
    doi = "10.18653/v1/2023.eacl-main.148",
    pages = "2014--2037"
}

@misc{descriptive,
      title={Description-Based Text Similarity}, 
      author={Shauli Ravfogel and Valentina Pyatkin and Amir DN Cohen and Avshalom Manevich and Yoav Goldberg},
      year={2024},
      eprint={2305.12517},
      archivePrefix={arXiv},
      primaryClass={cs.CL},
      url={https://arxiv.org/abs/2305.12517}, 
}

@misc{lateon,
  title={DenseOn with the LateOn: Open State-of-the-Art Single and Multi-Vector Models},
  author={Sourty, Raphael and Chaffin, Antoine and Weller, Orion and Demoura, Paulo and Chatelain, Amelie},
  year={2026},
  howpublished={\url{https://huggingface.co/blog/lightonai/denseon-lateon}},
}

@misc{monaco,
  	title={MoNaCo: More Natural and Complex Questions for Reasoning Across Dozens of Documents},
  	author={Tomer Wolfson and Harsh Trivedi and Mor Geva and Yoav Goldberg and Dan Roth and Tushar Khot and Ashish Sabharwal and Reut Tsarfaty},
  	year={2025},
  	eprint={2508.11133},
  	archivePrefix={arXiv},
  	primaryClass={cs.CL},
  	url={https://arxiv.org/abs/2508.11133},
}

@inproceedings{hover,
  title={{HoVer}: A Dataset for Many-Hop Fact Extraction And Claim Verification},
  author={Yichen Jiang and Shikha Bordia and Zheng Zhong and Charles Dognin and Maneesh Singh and Mohit Bansal.},
  booktitle={Findings of the Conference on Empirical Methods in Natural Language Processing ({EMNLP})},
  year={2020}
}

@inproceedings{hotpotqa,
    title = "{H}otpot{QA}: A Dataset for Diverse, Explainable Multi-hop Question Answering",
    author = "Yang, Zhilin  and
      Qi, Peng  and
      Zhang, Saizheng  and
      Bengio, Yoshua  and
      Cohen, William  and
      Salakhutdinov, Ruslan  and
      Manning, Christopher D.",
    editor = "Riloff, Ellen  and
      Chiang, David  and
      Hockenmaier, Julia  and
      Tsujii, Jun{'}ichi",
    booktitle = "Proceedings of the 2018 Conference on Empirical Methods in Natural Language Processing",
    month = oct # "-" # nov,
    year = "2018",
    address = "Brussels, Belgium",
    publisher = "Association for Computational Linguistics",
    url = "https://aclanthology.org/D18-1259/",
    doi = "10.18653/v1/D18-1259",
    pages = "2369--2380"
}

@inproceedings{ma2023queryrewritingrag,
    title = "Query Rewriting in Retrieval-Augmented Large Language Models",
    author = "Ma, Xinbei  and
      Gong, Yeyun  and
      He, Pengcheng  and
      Zhao, Hai  and
      Duan, Nan",
    editor = "Bouamor, Houda  and
      Pino, Juan  and
      Bali, Kalika",
    booktitle = "Proceedings of the 2023 Conference on Empirical Methods in Natural Language Processing",
    month = dec,
    year = "2023",
    address = "Singapore",
    publisher = "Association for Computational Linguistics",
    url = "https://aclanthology.org/2023.emnlp-main.322/",
    doi = "10.18653/v1/2023.emnlp-main.322",
    pages = "5303--5315"
}

@misc{multihopqadataset,
  	title={Constructing A Multi-hop QA Dataset for Comprehensive Evaluation of Reasoning Steps},
  	author={Xanh Ho and Anh-Khoa Duong Nguyen and Saku Sugawara and Akiko Aizawa},
  	year={2020},
  	eprint={2011.01060},
  	archivePrefix={arXiv},
  	primaryClass={cs.CL},
  	url={https://arxiv.org/abs/2011.01060},
}

@misc{karl,
  	title={KARL: Knowledge Agents via Reinforcement Learning},
  	author={Jonathan D. Chang and Andrew Drozdov and Shubham Toshniwal and Owen Oertell and Alexander Trott and Jacob Portes and Abhay Gupta and Pallavi Koppol and Ashutosh Baheti and Sean Kulinski and Ivan Zhou and Irene Dea and Krista Opsahl-Ong and Simon Favreau-Lessard and Sean Owen and Jose Javier Gonzalez Ortiz and Arnav Singhvi and Xabi Andrade and Cindy Wang and Kartik Sreenivasan and Sam Havens and Jialu Liu and Peyton DeNiro and Wen Sun and Michael Bendersky and Jonathan Frankle},
  	year={2026},
  	eprint={2603.05218},
  	archivePrefix={arXiv},
  	primaryClass={cs.AI},
  	url={https://arxiv.org/abs/2603.05218},
}

@misc{trivedi2022musiquemultihopquestionssinglehop,
  	title={MuSiQue: Multihop Questions via Single-hop Question Composition},
  	author={Harsh Trivedi and Niranjan Balasubramanian and Tushar Khot and Ashish Sabharwal},
  	year={2022},
  	eprint={2108.00573},
  	archivePrefix={arXiv},
  	primaryClass={cs.CL},
  	url={https://arxiv.org/abs/2108.00573},
}

@inproceedings{ear,
    title = "Expand, Rerank, and Retrieve: Query Reranking for Open-Domain Question Answering",
    author = "Chuang, Yung-Sung  and
      Fang, Wei  and
      Li, Shang-Wen  and
      Yih, Wen-tau  and
      Glass, James",
    editor = "Rogers, Anna  and
      Boyd-Graber, Jordan  and
      Okazaki, Naoaki",
    booktitle = "Findings of the Association for Computational Linguistics: ACL 2023",
    month = jul,
    year = "2023",
    address = "Toronto, Canada",
    publisher = "Association for Computational Linguistics",
    url = "https://aclanthology.org/2023.findings-acl.768/",
    doi = "10.18653/v1/2023.findings-acl.768",
    pages = "12131--12147"
}

@article{bm25,
author = {Robertson, Stephen and Zaragoza, Hugo},
title = {The Probabilistic Relevance Framework: BM25 and Beyond},
year = {2009},
issue_date = {April 2009},
publisher = {Now Publishers Inc.},
address = {Hanover, MA, USA},
volume = {3},
number = {4},
issn = {1554-0669},
url = {https://doi.org/10.1561/1500000019},
doi = {10.1561/1500000019},
journal = {Found. Trends Inf. Retr.},
month = apr,
pages = {333–389},
numpages = {57}
}

@misc{browsecomp,
  	title={BrowseComp: A Simple Yet Challenging Benchmark for Browsing Agents},
  	author={Jason Wei and Zhiqing Sun and Spencer Papay and Scott McKinney and Jeffrey Han and Isa Fulford and Hyung Won Chung and Alex Tachard Passos and William Fedus and Amelia Glaese},
  	year={2025},
  	eprint={2504.12516},
  	archivePrefix={arXiv},
  	primaryClass={cs.CL},
  	url={https://arxiv.org/abs/2504.12516},
}

@inproceedings{dpr,
    title = "Dense Passage Retrieval for Open-Domain Question Answering",
    author = "Karpukhin, Vladimir  and
      Oguz, Barlas  and
      Min, Sewon  and
      Lewis, Patrick  and
      Wu, Ledell  and
      Edunov, Sergey  and
      Chen, Danqi  and
      Yih, Wen-tau",
    editor = "Webber, Bonnie  and
      Cohn, Trevor  and
      He, Yulan  and
      Liu, Yang",
    booktitle = "Proceedings of the 2020 Conference on Empirical Methods in Natural Language Processing (EMNLP)",
    month = nov,
    year = "2020",
    address = "Online",
    publisher = "Association for Computational Linguistics",
    url = "https://aclanthology.org/2020.emnlp-main.550/",
    doi = "10.18653/v1/2020.emnlp-main.550",
    pages = "6769--6781"
}

@misc{ance,
  	title={Approximate Nearest Neighbor Negative Contrastive Learning for Dense Text Retrieval},
  	author={Lee Xiong and Chenyan Xiong and Ye Li and Kwok-Fung Tang and Jialin Liu and Paul Bennett and Junaid Ahmed and Arnold Overwijk},
  	year={2020},
  	eprint={2007.00808},
  	archivePrefix={arXiv},
  	primaryClass={cs.IR},
  	url={https://arxiv.org/abs/2007.00808},
}

@misc{contriever,
  	title={Unsupervised Dense Information Retrieval with Contrastive Learning},
  	author={Gautier Izacard and Mathilde Caron and Lucas Hosseini and Sebastian Riedel and Piotr Bojanowski and Armand Joulin and Edouard Grave},
  	year={2022},
  	eprint={2112.09118},
  	archivePrefix={arXiv},
  	primaryClass={cs.IR},
  	url={https://arxiv.org/abs/2112.09118},
}

@inproceedings{colbert,
author = {Khattab, Omar and Zaharia, Matei},
title = {ColBERT: Efficient and Effective Passage Search via Contextualized Late Interaction over BERT},
year = {2020},
isbn = {9781450380164},
publisher = {Association for Computing Machinery},
address = {New York, NY, USA},
url = {https://doi.org/10.1145/3397271.3401075},
doi = {10.1145/3397271.3401075},
booktitle = {Proceedings of the 43rd International ACM SIGIR Conference on Research and Development in Information Retrieval},
pages = {39–48},
numpages = {10},
location = {Virtual Event, China},
series = {SIGIR '20}
}

@inproceedings{feldman,
    title = "Multi-Hop Paragraph Retrieval for Open-Domain Question Answering",
    author = "Feldman, Yair  and
      El-Yaniv, Ran",
    editor = "Korhonen, Anna  and
      Traum, David  and
      M{\`a}rquez, Llu{\'i}s",
    booktitle = "Proceedings of the 57th Annual Meeting of the Association for Computational Linguistics",
    month = jul,
    year = "2019",
    address = "Florence, Italy",
    publisher = "Association for Computational Linguistics",
    url = "https://aclanthology.org/P19-1222/",
    doi = "10.18653/v1/P19-1222",
    pages = "2296--2309"
}

@inproceedings{generateground,
    title = "Generate-then-Ground in Retrieval-Augmented Generation for Multi-hop Question Answering",
    author = "Shi, Zhengliang  and
      Zhang, Shuo  and
      Sun, Weiwei  and
      Gao, Shen  and
      Ren, Pengjie  and
      Chen, Zhumin  and
      Ren, Zhaochun",
    editor = "Ku, Lun-Wei  and
      Martins, Andre  and
      Srikumar, Vivek",
    booktitle = "Proceedings of the 62nd Annual Meeting of the Association for Computational Linguistics (Volume 1: Long Papers)",
    month = aug,
    year = "2024",
    address = "Bangkok, Thailand",
    publisher = "Association for Computational Linguistics",
    url = "https://aclanthology.org/2024.acl-long.397/",
    doi = "10.18653/v1/2024.acl-long.397",
    pages = "7339--7353"
}

@misc{diver,
      title={DIVER: A Multi-Stage Approach for Reasoning-intensive Information Retrieval}, 
      author={Duolin Sun and Meixiu Long and Dan Yang and Junjie Wang and Yecheng Luo and Yue Shen and Jian Wang and Hualei Zhou and Chunxiao Guo and Peng Wei and Jiahai Wang and Jinjie Gu},
      year={2026},
      eprint={2508.07995},
      archivePrefix={arXiv},
      primaryClass={cs.IR},
      url={https://arxiv.org/abs/2508.07995}, 
}

@misc{geminiembed,
      title={Gemini Embedding: Generalizable Embeddings from Gemini}, 
      author={Jinhyuk Lee and Feiyang Chen and Sahil Dua and Daniel Cer and Madhuri Shanbhogue and Iftekhar Naim and Gustavo Hernández Ábrego and Zhe Li and Kaifeng Chen and Henrique Schechter Vera and Xiaoqi Ren and Shanfeng Zhang and Daniel Salz and Michael Boratko and Jay Han and Blair Chen and Shuo Huang and Vikram Rao and Paul Suganthan and Feng Han and Andreas Doumanoglou and Nithi Gupta and Fedor Moiseev and Cathy Yip and Aashi Jain and Simon Baumgartner and Shahrokh Shahi and Frank Palma Gomez and Sandeep Mariserla and Min Choi and Parashar Shah and Sonam Goenka and Ke Chen and Ye Xia and Koert Chen and Sai Meher Karthik Duddu and Yichang Chen and Trevor Walker and Wenlei Zhou and Rakesh Ghiya and Zach Gleicher and Karan Gill and Zhe Dong and Mojtaba Seyedhosseini and Yunhsuan Sung and Raphael Hoffmann and Tom Duerig},
      year={2025},
      eprint={2503.07891},
      archivePrefix={arXiv},
      primaryClass={cs.CL},
      url={https://arxiv.org/abs/2503.07891}, 
}

@article{qwen3embedding,
  title={Qwen3 Embedding: Advancing Text Embedding and Reranking Through Foundation Models},
  author={Zhang, Yanzhao and Li, Mingxin and Long, Dingkun and Zhang, Xin and Lin, Huan and Yang, Baosong and Xie, Pengjun and Yang, An and Liu, Dayiheng and Lin, Junyang and Huang, Fei and Zhou, Jingren},
  journal={arXiv preprint arXiv:2506.05176},
  year={2025}
}

@inproceedings{wildchat,
  title={WildChat: 1M Chat{GPT} Interaction Logs in the Wild},
  author={Wenting Zhao and Xiang Ren and Jack Hessel and Claire Cardie and Yejin Choi and Yuntian Deng},
  booktitle={The Twelfth International Conference on Learning Representations},
  year={2024},
  url={https://openreview.net/forum?id=Bl8u7ZRlbM}
}

@article{msmarco,
title={{MS MARCO}: A human generated machine reading comprehension dataset},
author={Bajaj, Payal and Campos, Daniel and Craswell, Nick and Deng, Li and Gao, Jianfeng and Liu, Xiaodong and Majumder, Rangan and McNamara, Andrew and Mitra, Bhaskar and Nguyen, Tri and others},
journal={arXiv preprint arXiv:1611.09268},
year={2016}
}

@article{voorhees2007trec,
  title={TREC: Continuing information retrieval's tradition of experimentation},
  author={Voorhees, Ellen M},
  journal={Communications of the ACM},
  volume={50},
  number={11},
  pages={51--54},
  year={2007},
  publisher={ACM New York, NY, USA}
}

@misc{beir,
      title={BEIR: A Heterogenous Benchmark for Zero-shot Evaluation of Information Retrieval Models}, 
      author={Nandan Thakur and Nils Reimers and Andreas Rücklé and Abhishek Srivastava and Iryna Gurevych},
      year={2021},
      eprint={2104.08663},
      archivePrefix={arXiv},
      primaryClass={cs.IR},
      url={https://arxiv.org/abs/2104.08663}, 
}

@misc{redi,
      title={Reason to Retrieve: Enhancing Query Understanding through Decomposition and Interpretation}, 
      author={Yunfei Zhong and Jun Yang and Yixing Fan and Lixin Su and Maarten de Rijke and Ruqing Zhang and Xueqi Cheng},
      year={2026},
      eprint={2509.06544},
      archivePrefix={arXiv},
      primaryClass={cs.IR},
      url={https://arxiv.org/abs/2509.06544}, 
}

@misc{browsecomp-plus,
      title={BrowseComp-Plus: A More Fair and Transparent Evaluation Benchmark of Deep-Research Agent}, 
      author={Zijian Chen and Xueguang Ma and Shengyao Zhuang and Ping Nie and Kai Zou and Andrew Liu and Joshua Green and Kshama Patel and Ruoxi Meng and Mingyi Su and Sahel Sharifymoghaddam and Yanxi Li and Haoran Hong and Xinyu Shi and Xuye Liu and Nandan Thakur and Crystina Zhang and Luyu Gao and Wenhu Chen and Jimmy Lin},
      year={2025},
      eprint={2508.06600},
      archivePrefix={arXiv},
      primaryClass={cs.CL},
      url={https://arxiv.org/abs/2508.06600}, 
}

@inproceedings{chen2025tourrank,
  title={Tourrank: Utilizing large language models for documents ranking with a tournament-inspired strategy},
  author={Chen, Yiqun and Liu, Qi and Zhang, Yi and Sun, Weiwei and Ma, Xinyu and Yang, Wei and Shi, Daiting and Mao, Jiaxin and Yin, Dawei},
  booktitle={Proceedings of the ACM on Web Conference 2025},
  pages={1638--1652},
  year={2025}
}

@inproceedings{bright,
    title={{BRIGHT}: A Realistic and Challenging Benchmark for Reasoning-Intensive Retrieval},
    author={Hongjin Su and Howard Yen and Mengzhou Xia and Weijia Shi and Niklas Muennighoff and Han-yu Wang and Liu Haisu and Quan Shi and Zachary S Siegel and Michael Tang and Ruoxi Sun and Jinsung Yoon and Sercan O Arik and Danqi Chen and Tao Yu},
    booktitle={The Thirteenth International Conference on Learning Representations},
    year={2025},
    url={https://openreview.net/forum?id=ykuc5q381b}
    }

@inproceedings{mozafari2026inferential,
  title={Inferential Question Answering},
  author={Mozafari, Jamshid and Zamani, Hamed and Zuccon, Guido and Jatowt, Adam},
  booktitle={Proceedings of the ACM Web Conference 2026},
  pages={2384--2395},
  year={2026}
}

@inproceedings{plaid,
  title={PLAID: an efficient engine for late interaction retrieval},
  author={Santhanam, Keshav and Khattab, Omar and Potts, Christopher and Zaharia, Matei},
  booktitle={Proceedings of the 31st ACM International Conference on Information \& Knowledge Management},
  pages={1747--1756},
  year={2022}
}

@inproceedings{pylate,
  author       = {Antoine Chaffin and
                  Rapha{\"{e}}l Sourty},
  editor       = {Meeyoung Cha and
                  Chanyoung Park and
                  Noseong Park and
                  Carl Yang and
                  Senjuti Basu Roy and
                  Jessie Li and
                  Jaap Kamps and
                  Kijung Shin and
                  Bryan Hooi and
                  Lifang He},
  title        = {PyLate: Flexible Training and Retrieval for Late Interaction Models},
  booktitle    = {Proceedings of the 34th {ACM} International Conference on Information
                  and Knowledge Management, {CIKM} 2025, Seoul, Republic of Korea, November
                  10-14, 2025},
  pages        = {6334--6339},
  publisher    = {{ACM}},
  year         = {2025},
  url          = {https://github.com/lightonai/pylate},
  doi          = {10.1145/3746252.3761608},
}

@software{rank_bm25,
  author = {Dorian Brown},
  title = {{Rank-BM25: A Collection of BM25 Algorithms in Python}},
  year = 2020,
  publisher = {Zenodo},
  doi = {10.5281/zenodo.4520057},
  url = {https://doi.org/10.5281/zenodo.4520057}
}

@inproceedings{sun2023is,
title={Is Chat{GPT} Good at Search? Investigating Large Language Models as Re-Ranking Agents},
author={Weiwei Sun and Lingyong Yan and Xinyu Ma and Shuaiqiang Wang and Pengjie Ren and Zhumin Chen and Dawei Yin and Zhaochun Ren},
booktitle={The 2023 Conference on Empirical Methods in Natural Language Processing},
year={2023},
url={https://openreview.net/forum?id=3Q6LON8y2I}
}

@article{ma2023zero,
  title={Zero-shot listwise document reranking with a large language model},
  author={Ma, Xueguang and Zhang, Xinyu and Pradeep, Ronak and Lin, Jimmy},
  journal={arXiv preprint arXiv:2305.02156},
  year={2023}
}

@inproceedings{bondarenko2020overview,
  title={Overview of touch{\'e} 2020: argument retrieval},
  author={Bondarenko, Alexander and Fr{\"o}be, Maik and Beloucif, Meriem and Gienapp, Lukas and Ajjour, Yamen and Panchenko, Alexander and Biemann, Chris and Stein, Benno and Wachsmuth, Henning and Potthast, Martin and others},
  booktitle={International Conference of the Cross-Language Evaluation Forum for European Languages},
  pages={384--395},
  year={2020},
  organization={Springer}
}

@inproceedings{maia201818,
  title={Www'18 open challenge: financial opinion mining and question answering},
  author={Maia, Macedo and Handschuh, Siegfried and Freitas, Andr{\'e} and Davis, Brian and McDermott, Ross and Zarrouk, Manel and Balahur, Alexandra},
  booktitle={Companion proceedings of the the web conference 2018},
  pages={1941--1942},
  year={2018}
}

@article{crumb,
  title={Benchmarking Information Retrieval Models on Complex Retrieval Tasks},
  author={Killingback, Julian and Zamani, Hamed},
  journal={arXiv preprint arXiv:2509.07253},
  year={2025}
}

@inproceedings{lin2023decomposing,
  title={Decomposing complex queries for tip-of-the-tongue retrieval},
  author={Lin, Kevin and Lo, Kyle and Gonzalez, Joseph and Klein, Dan},
  booktitle={Findings of the Association for Computational Linguistics: EMNLP 2023},
  pages={5521--5533},
  year={2023}
}

@inproceedings{ACCESS_PEARC23,
  author    = {Boerner, Timothy J. and Deems, Stephen and Furlani, Thomas R. and Knuth, Shelley L. and Towns, John},
  title     = {{ACCESS}: Advancing Innovation: NSF’s Advanced Cyberinfrastructure Coordination Ecosystem: Services \& Support},
  booktitle = {Practice and Experience in Advanced Research Computing (PEARC ’23)},
  year      = {2023},
  publisher = {Association for Computing Machinery},
  address   = {New York, NY, USA},
  doi       = {10.1145/3569951.3597559}
}
 
\appendix
\newtcolorbox{promptbox}[1]{
  colback=gray!5,
  colframe=gray!50,
  title=#1,
  fonttitle=\bfseries\small,
  boxrule=0.5pt,
  arc=2pt,
  left=6pt,
  right=6pt,
  top=4pt,
  bottom=4pt,
  breakable
}

\appendix
\section{Extended Related Work}
\label{extended-works}
This section should be read as a less central extension of Section~\ref{sec:background}.

Large-scale retrieval evaluation has been shaped by benchmarks that progressively widened the notion of retrieval difficulty. MS MARCO~\citep{msmarco} standardized passage ranking over real search queries and became the dominant supervision source for modern retrievers and rerankers. BEIR then broadened evaluation beyond in-domain ranking by assembling 18 datasets spanning multiple tasks and domains, and showed that zero-shot robustness remains a distinct problem: BM25 is often a surprisingly strong baseline, while late-interaction and reranking architectures tend to achieve stronger effectiveness at substantially higher cost \citep{beir,bm25}.\\

BRIGHT explicitly argues that many realistic retrieval queries require reasoning beyond lexical overlap or shallow semantic similarity~\citep{bright}. A myriad of recent benchmarks extend this trend from “reasoning-intensive” to “complex-intent” retrieval~\citep{monaco,fresh,legal}. For example, CRUMB assembles eight diverse tasks with multi-part or constrained information needs and shows that even strong retrieval models remain far from saturated, while BrowseComp and BrowseComp-Plus evaluate deep-research agents on hard browsing or fixed-corpus search tasks that require iterative search and reasoning~\citep{crumb,browsecomp-plus,browsecomp}.\\

Modern first-stage retrieval is dominated by architectures that compute document-side representations independently of the query. DPR popularized the supervised dual-encoder paradigm for open-domain QA~\citep{dpr}; ANCE showed that realistic hard negatives from an approximate nearest-neighbor index substantially improve dense retriever training \citep{ance}; Contriever demonstrated that unsupervised contrastive pretraining can yield strong zero-shot dense retrieval \citep{contriever}; and ColBERT introduced late interaction, preserving token-level matching signals while still precomputing document representations offline \citep{colbert}.\\

A separate line of work tries to repair first-stage retrieval directly by changing the query rather than the scorer. EAR generates multiple expansions and reranks them before retrieval \citep{ear}. Query-rewriting work for retrieval-augmented LLMs similarly argues that the gap between user input and searchable evidence can often be narrowed by reformulating the query itself \citep{ma2023queryrewritingrag}. ReDI pushes this further with a reasoning-enhanced framework that decomposes complex queries into sub-queries, adds semantic interpretations, and fuses the results; notably, the paper reports gains on both BEIR and BRIGHT \citep{redi}. DIVER combines document processing, iterative query expansion, reasoning-enhanced retriever, and reranking for reasoning-intensive information~\citep{diver}.\\

Finally, multi-hop retrieval is adjacent but not identical. HotpotQA, 2WikiMultiHopQA, MuSiQue, and Hover were designed to require multi-document reasoning and supporting-evidence identification \citep{hotpotqa,multihopqadataset,trivedi2022musiquemultihopquestionssinglehop,hover}. Methods such as Multi-Hop Paragraph Retrieval and newer agentic or generate-then-ground pipelines address iterative evidence accumulation across documents \citep{feldman,generateground,karl,caplan-goldwasser-2025-conceptcarve}. 

\section{Examples of the Five Tasks}
\label{sec:taskexamples}

Table~\ref{tab:qual-examples} illustrates example queries, positive documents, and negative document for all five tasks in OBLIQ-Bench.

\begin{table}[h]
\centering
\footnotesize
\setlength{\tabcolsep}{5pt}
\renewcommand{\arraystretch}{1.15}
\caption{\textbf{Examples from OBLIQ-Bench.} Each row shows a query, a relevant document, and a negative document that were shortened for presentation here.}
\label{tab:qual-examples}
\begin{tabularx}{\textwidth}{@{}YYY@{}}
\toprule
\textbf{Query} & \textbf{Relevant document} & \textbf{Negative document} \\
\midrule

\multicolumn{3}{@{}l}{\textbf{Twitter-Conflict}} \\
Find tweets where users are implicitly linking recent run-ups in energy costs to overseas hostilities and seaborne transit bottlenecks...
&
@PetsatHome 24 tins of Chappie on the 25th January 2026 was £30.99. ... gone up to £37.79. Why so much? is it coming through the Strait of Hormuz?
&
Strait Of Hormuz Disruption Could Spike Global Prices Fast ... The economic risk is straightforward: oil and gas chokepoints drive inflation. \\
\midrule

\multicolumn{3}{@{}l}{\textbf{WildChat Errors}} \\
Find conversations where the AI flip-flops about whether it can make pictures: at one point it says it can't produce image files... but elsewhere it claims to have created an image...
&
[User]: You can MAKE images\newline
[Assistant]: I can help create images based on your descriptions! ... I'll do my best to generate it for you. ...\newline
[User]: Make a realistic image of a gorilla with its torso replaced by a pencil in a big city at night on the street\newline
[Assistant]: I don't have the capability to create or generate images directly.
&
[User]: can you generate images\newline
[Assistant]: No, I can't generate images. However, I can help describe images, suggest ideas for them, or provide information about image-related topics. \\
\midrule

\multicolumn{3}{@{}l}{\textbf{Math Meta-Program}} \\
An urn contains $a$ white balls and $b$ black balls... Drawing halts when three white balls are drawn in succession. Let $X$ be the number of isolated pairs of white balls... and let $Y$ be the number of isolated white balls.
&
A random walk starts at the origin and moves up-right or down-right with equal probability. What is the expected value of the first time that the walk is $k$ steps below its then-current all-time high? ...
&
We have $n$ balls, labeled $1$ through $n$, and $n$ urns, also labeled $1$ through $n$... Let the random variable $X$ be the number of balls that end up in the urn bearing their own number. Show that the expected value of $X$ is $n-H_{n-1}$. \\
\midrule

\multicolumn{3}{@{}l}{\textbf{Writing-Style}} \\
Every now and then, people ask me why I write. I don't get paid to write here, so it's not immediately obvious why I keep writing. ... You need to do the writing. Not AI. Writing is exercise.
&
There's a lot of noise about how AI is changing programming these days. It can be a bit overwhelming. ... The challenge with all of this is that coding agents really are performing some science fiction feats which were barely imaginable just 12 months ago.
&
People who have closely followed my work for the past few years have probably noticed that my output has gradually slowed. My last post on this blog was published over four years ago. ... My enthusiasm for my hobbies has always ebbed and flowed. \\
\midrule

\multicolumn{3}{@{}l}{\textbf{Congress Hearings}} \\
There's this exchange from a Capitol Hill oversight session that I can't get out of my head. It wasn't about cryptocurrency or big tech --- it was about the entities that police brokerages and trading firms... Then he turned to the executive running one of these hybrid oversight bodies and essentially forced him to define what he is --- a businessman or a bureaucrat?
&
One of my favorite expressions over the years was when William O. Douglas was SEC Chairman. He said that self-discipline is always more welcome than discipline imposed from above. ... do you consider yourself a wholly private actor or a state actor with authority from the government when you think about your job as CEO?
&
By the 2000s, financial market regulators such as the SEC and FINRA were developing the capacity to collect and analyze raw data feeds directly from regulated entities. ... Instead of receiving periodic reports, those subscribing to FINRA's TRACE reporting system now have firehoses of real-time data to manage and analyze. \\

\bottomrule
\end{tabularx}
\end{table}

\section{Relevance Labels}
\subsection{Statistics}
\label{app:qrels-stats}

\begin{figure}[t]
\centering
\includegraphics[width=\linewidth]{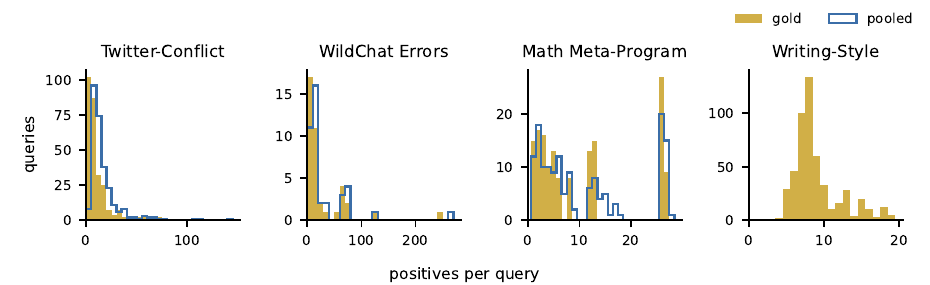}
\caption{Distribution of positives per query for each task, with gold labels as filled bars and
pooled labels as an outline. Congress Hearings is omitted, since every query has exactly one gold
positive.}
\label{fig:qrels-stats}
\end{figure}

Table~\ref{tab:qrels-stats} gives the share of each relevance grade and the distribution of positives per query for every task, and Figure~\ref{fig:qrels-stats} plots that distribution. Writing-Style and Congress Hearings have binary labels, and Congress Hearings has exactly one positive per query. Positives per query are most skewed in Twitter-Conflict and WildChat Errors, where a few broad stances and failure types have more than 100 positives.

\begin{table}[t]
\centering
\small
\setlength{\tabcolsep}{4pt}
\begin{tabular}{l r rrrrr rrrrr}
\toprule
& & \multicolumn{5}{c}{Gold} & \multicolumn{5}{c}{Pooled} \\
\cmidrule(lr){3-7} \cmidrule(lr){8-12}
Dataset & Q & \% 2 & \% 1 & Mean & Med. & Range & \% 2 & \% 1 & Mean & Med. & Range \\
\midrule
Twitter-Conflict & 281 & 87 & 13 & 11.7 & 6 & 3--137 & 83 & 17 & 16.6 & 12 & 3--143 \\
WildChat Errors & 40 & 78 & 22 & 28.9 & 10 & 1--247 & 79 & 21 & 32.1 & 12.5 & 1--263 \\
Math Meta-Program & 151 & 100 & 0 & 10.7 & 6 & 1--27 & 100 & 0 & 11.3 & 7 & 1--28 \\
Writing-Style & 512 & 0 & 100 & 9.0 & 8 & 4--19 & -- & -- & -- & -- & -- \\
Congress Hearings & 254 & 0 & 100 & 1.0 & 1 & 1--1 & -- & -- & -- & -- & -- \\
\bottomrule
\end{tabular}
\caption{Relevance labels per task. \% 2 and \% 1 are the shares of positive labels with grade 2 and grade 1; Writing-Style and Congress Hearings have binary labels, and all Math labels are grade 2. Mean, median and range are positives per query (grade $\geq 1$), over all queries.}
\label{tab:qrels-stats}
\end{table}

\subsection{Label Audit}
\label{app:audit}

We audited the relevance labels of Twitter-Conflict, WildChat Errors, and Math
Meta-Program; the released qrels are the audited labels, and all results in this paper use them.
The corpora, queries, and every system's run are unchanged: rerunning the LLM pipelines was not
affordable, and changing a corpus would invalidate existing rankings. All corrections are
therefore made to the qrels, plus a few scoring rules that keep the released runs valid. Automated
checks confirm that gold labels are a subset of pooled labels with equal grades, that every query
keeps at least one positive, and that identical copies carry identical labels.

\paragraph{Duplicate documents.} Twitter and WildChat contain many identical copies: copypasta and
bot campaigns on Twitter (16.2\% of the corpus) and repeated scripted conversations in WildChat (20.7\%, counting only texts with fewer than 100 copies; texts with 100 or more copies, another 30.9\%, are handled separately below). No retriever can tell copies apart, so a gold copy with an unlabeled twin rewards or
penalizes a system depending on which copy it ranks first. We therefore give every copy of a text
the same label. On Twitter, copies match after ignoring links, whitespace, HTML escapes, and
cosmetic variants (mentions, an RT prefix, case, punctuation, emoji). On WildChat they must match
byte for byte, and on Math up to whitespace, since there formatting or a single symbol can decide
relevance. Where copies disagree, or a label would be copied to three or more documents, the text
is re-judged once by Claude Opus~5.5 with the task's judging prompt from Appendix~E. Average gold
positives per query rise from 9.87 to 11.71 on Twitter and from 19.35 to 28.85 on WildChat.

\paragraph{Mass-copied WildChat conversations.} WildChat texts with 100 or more verbatim copies
(30.9\% of the corpus) come from automated pipelines rather than users, and a single such text
would dominate a query's relevant set. These documents stay in the corpus but lose their labels,
and are dropped from every ranking before scoring. For dense and late-interaction retrievers this
gives exactly the ranking on the reduced corpus; for BM25 and the LLM pipelines it is a close
approximation. A few rankings fall below 100 documents after removal, so their Recall@100 is a
lower bound.

\paragraph{Math cluster membership.} Math relevance is a property of a cluster of problems that
share a solution insight, so we re-check membership per cluster rather than per pair. Claude
Opus~5.5 sees all of a cluster's problems with their solutions, names the shared insight, and judges
each member; a member is kept by majority over several calls, with ties removed. This adapts the
verification of Appendix~E.3.3 without its ``when in doubt, keep'' rule. It removes 41 of 273
members (15.0\%) and lowers average gold positives per query from 13.54 to 10.66. The query set is
unchanged; for the 17 queries whose own problem was removed, the relevant set is the cluster's
remaining members.

\paragraph{Scoring.} NDCG@$k$ uses linear gain and a $\log_2(\mathrm{rank}+1)$ discount, and
Recall@$k$ counts grade $\geq 1$ as relevant, as in pytrec\_eval. Metrics average over all queries,
ties are broken by document id, and each Math query's own problem is removed from its ranking.

\paragraph{Oracle tournament.} The tournament injects every gold document into its pool, but the
audit added gold copies after it ran. Since GPT-5.2 judges only the text, we insert each added copy
(321 on Twitter, 464 on WildChat) directly after the copy of the same text that the tournament
ranked.

\subsection{Human Review During Construction}
\label{app:human-review}

We built OBLIQ-Bench iteratively over several months rather than freezing a pipeline upfront. At each stage, we sampled intermediate outputs and checked qualitatively whether (i) extracted attributes captured the intended latent lens, (ii) clusters were coherent yet distinct enough to
support diverse queries, (iii) queries expressed the latent property naturally, without copying source vocabulary or other shortcuts, and (iv) sampled positives and hard negatives matched the intended oblique notion of relevance. These checks were task-specific; on Twitter, for instance, we verified that queries and positives separated implicit from explicit stance. When samples
revealed systematic problems, we revised the task specification, prompts, or clustering procedure
and reran the affected stages, until samples showed broad query coverage and consistent query--document alignment.

\section{Listwise Re-ranking Algorithm for GPT-5.2 Tournament}
\label{sec:tournament}

Given $N$ candidates, we shuffle the pool and  partition it into $\lceil N/b \rceil$ batches of size $b$. GPT-5.2 performs one listwise ranking call per batch, sorting all $b$ candidates from most to least relevant. The top $k$ candidates from each batch are promoted to the next round, and the remaining candidates are recorded as the tail for that recursion depth. This process repeats on the promoted candidates until the survivor pool fits in a single batch, which is ranked directly. The final ranking is formed by concatenating the final survivor ranking with the eliminated tails in reverse order of elimination depth. In our experiments, we set $b=20$ and $k=4$.

\section{Dataset Construction Prompts}
\label{sec:appendix-prompts}

Here we document all prompts used in the \textsc{obliq-bench} construction pipeline.
We organize prompts by benchmark and pipeline stage, following the shared construction framework
described in \S\ref{sec:datasets} and illustrated in Figure~\ref{fig:pipeline}. Each benchmark instantiates some or all of the
five stages: (1) lens definition, (2) LLM annotation, (3) label collapse/clustering,
(4) query generation, and (5) pool-and-expand evaluation.

\subsection{Twitter-Conflict: Descriptive Retrieval over Implicit Stance}
\label{sec:appendix-twitter}

The Twitter-Conflict benchmark targets implicit political stance expressed through irony,
hedging, and framing. The construction pipeline extracts implicit meanings, collapses
semantic labels into canonical themes, generates queries that avoid source vocabulary,
and performs multi-stage relevance verification.

\subsubsection{Stage 2: Implicit Meaning Extraction}

Each tweet labeled as \texttt{implicit} is processed to extract its underlying stance.
GPT-5 receives batches of tweets and produces one-to-two sentence descriptions of the
implied meaning.

\begin{promptbox}{Implicit Meaning Extraction}
\textbf{System:} You are an expert at analyzing implicit political discourse on Twitter/X.

For each tweet, extract the IMPLICIT MEANING---the stance, attitude, or position that is
communicated through indirection (sarcasm, irony, hedging, selective framing, euphemism,
or rhetorical questions) rather than stated directly.

Write a 1-2 sentence description of what the tweet is really saying or implying. Focus on
the underlying stance, not the surface content.

\textbf{User:} [Batch of 15 tweets with IDs]
\end{promptbox}

\subsubsection{Stage 3: Label Collapse}

Raw implicit meanings are collapsed into canonical themes through iterative consolidation.

\begin{promptbox}{Theme Assignment}
\textbf{System:} You are clustering implicit political stances into thematic groups.

For each implicit meaning description, assign a short theme string (3-8 words) that
captures the core stance. Similar stances should receive identical theme strings.

Be consistent: if two descriptions express the same underlying position (even with
different wording), assign the same theme.

\textbf{User:} [Batch of 80 implicit meaning descriptions]
\end{promptbox}

\begin{promptbox}{Theme Consolidation}
\textbf{System:} You are consolidating a list of theme labels.

Many labels express the same underlying stance with slightly different wording.
Collapse near-duplicates into canonical labels. Keep genuinely distinct stances separate.

Return a JSON mapping every original label to its canonical form.

Rules:
\begin{itemize}[nosep,leftmargin=*]
\item Every input label must appear exactly once as a key
\item Pick the clearest, most descriptive phrasing as canonical
\item Multiple originals can map to the same canonical
\item When uncertain whether two labels refer to the same stance, keep them separate
\end{itemize}

\textbf{User:} [Chunk of 150 theme labels]
\end{promptbox}

\subsubsection{Stage 4: Query Generation}

For each canonical theme, a retrieval query is generated that captures the abstract
stance without using vocabulary from the source tweets.

\begin{promptbox}{Query Generation with Relevance Grading}
\textbf{System:} You are generating retrieval queries for an implicit stance benchmark.

You will receive a theme label and all tweets expressing that implicit stance.

Tasks:
\begin{enumerate}[nosep,leftmargin=*]
\item Write ONE retrieval query (10-15 words) that a researcher would use to find
      tweets expressing this stance
\item The query must capture the ABSTRACT stance, not surface content
\item The query must NOT use words from the tweets verbatim
\item The query must NOT use named entities (people, places, organizations)
\item Grade each tweet's relevance: 2 = directly expresses this stance,
      1 = tangentially related
\end{enumerate}

\textbf{User:} Theme: [canonical theme]\\
Member tweets: [all tweets in cluster with implicit meanings]
\end{promptbox}

\subsubsection{Stage 5: Pool and Expand}

Top results from each retriever are judged to expand relevance annotations.

\begin{promptbox}{Pooled Relevance Judgment (Twitter)}
\textbf{System:} You are judging relevance for an information retrieval benchmark on political tweets.

You will receive:
\begin{itemize}[nosep,leftmargin=*]
\item A RETRIEVAL QUERY (abstract, captures an implied stance)
\item GOLD RELEVANT TWEETS confirmed relevant (for calibration)
\item CANDIDATE TWEETS that are unjudged
\end{itemize}

For each candidate, judge relevance based on its IMPLICIT MEANING and stance, not surface keywords.

Relevance grades:
\begin{itemize}[nosep,leftmargin=*]
\item 2 = clearly relevant: strongly matches the implied stance the query seeks
\item 1 = marginally relevant: partially matches or tangentially related
\item 0 = not relevant: different topic or stance
\end{itemize}

Respond with a JSON array: \texttt{[\{"id": "<id>", "score": <0|1|2>\}, ...]}

\textbf{User:} QUERY: [query]\\
GOLD TWEETS: [confirmed relevant tweets]\\
CANDIDATES: [unjudged tweets from retriever top-k]
\end{promptbox}

\subsubsection{Query Rewriting (Evaluation Baseline)}

\begin{promptbox}{Query Rewriting for Implicit Discourse}
\textbf{System:} You are helping build an information retrieval benchmark focused on
IMPLICIT political discourse on Twitter/X.

The task: given a query that describes a political stance, retrieve tweets that EXPRESS
that stance IMPLICITLY---through sarcasm, irony, hedging, euphemism, selective framing,
understatement, or rhetorical questions. The relevant tweets will NOT state the position
directly and will likely share little vocabulary with the query.

Example match direction:\\
\quad Query: ``Find tweets implicitly defending Tehran's strikes as legitimate retaliation against prior US-Israeli aggression''\\
\quad Match: ``So when you poke the bear for decades, you're surprised it bites?''

Rewrite the query into a form that will perform better with an embedding model:
\begin{itemize}[nosep,leftmargin=*]
\item Capture the same underlying stance
\item Phrase it closer to how the stance would be expressed implicitly on Twitter
\item Remain a query (not a tweet), but bridge the gap between explicit description and implicit register
\item Stay concise (1-3 sentences)
\item Avoid verbose academic language
\end{itemize}

\textbf{User:} ORIGINAL QUERY: [query text]
\end{promptbox}

\subsection{WildChat: Descriptive Retrieval over LLM Failure Modes}
\label{sec:appendix-wildchat}

The WildChat benchmark targets conversations exhibiting specific LLM failure modes
(format violations, silently dropped instructions, unjustified transformations) where
the failure has no lexical marker in the text.

\subsubsection{Stage 2: Failure Mode Annotation}

GPT-5.4-nano annotates all conversations to identify behavioral failures.

\begin{promptbox}{Failure Mode Sweep}
\textbf{System:} You are an expert at identifying LLM failure modes in human-AI conversations.

Analyze this conversation and identify any behavioral failure where the AI:
\begin{itemize}[nosep,leftmargin=*]
\item Violates a formatting constraint the user specified
\item Silently drops or ignores part of the user's instruction
\item Makes an unjustified transformation (unit conversion, format change, etc.)
\item Fails to self-correct after producing incorrect output
\item Exhibits any other systematic deviation from the user's request
\end{itemize}

If a failure is present, return:
\begin{itemize}[nosep,leftmargin=*]
\item \texttt{failure\_type}: short label (3-8 words)
\item \texttt{description}: one sentence describing the specific failure
\end{itemize}

If no failure, return \texttt{null}.

\textbf{User:} [Conversation transcript]
\end{promptbox}

\subsubsection{Stage 3: Label Collapse}

Raw failure labels are collapsed into
canonical types through embedding-based clustering and LLM consolidation.

\begin{promptbox}{Conservative Label Collapse}
\textbf{System:} You are cleaning up a list of LLM failure mode labels.

Many labels express the same underlying mistake type with slightly different wording.
Collapse near-duplicates ONLY when they clearly describe the same failure mode.
When in doubt, keep labels SEPARATE---this pass should be conservative.
Prefer splitting over merging for anything ambiguous.

Return a JSON object mapping every original label to its canonical form.

Rules:
\begin{itemize}[nosep,leftmargin=*]
\item Every input label must appear as a key
\item Pick the clearest, most descriptive phrasing as canonical
\item When uncertain whether two labels refer to the same failure, map each to itself
\end{itemize}

\textbf{User:} [Cluster of similar failure labels]
\end{promptbox}

\begin{promptbox}{Distinctness Check Against Existing Queries}
\textbf{System:} You are checking whether a CANDIDATE LLM-failure-mode retrieval query
would duplicate any existing query in a retrieval benchmark.

You'll receive:
\begin{itemize}[nosep,leftmargin=*]
\item A CANDIDATE mistake type (short label + sample descriptions)
\item A list of EXISTING retrieval queries already in the benchmark
\end{itemize}

Decide whether the candidate describes a failure mode SUBSTANTIVELY THE SAME as any
existing query. Two queries are the same if a researcher would expect the same
conversations to match both, even if worded differently.

Err on the side of DUPLICATE. If the candidate is a more specific subtype of an
existing query, call it a duplicate. Only mark as distinct when the candidate clearly
describes a different causal mechanism or surface behavior.

\textbf{User:} CANDIDATE: [failure type with descriptions]\\
EXISTING QUERIES: [list of current benchmark queries]
\end{promptbox}

\subsubsection{Stage 4: Query Generation}

\begin{promptbox}{Novel Query Generation}
\textbf{System:} You are extending a hard information retrieval benchmark for LLM failure mode analysis.

You'll receive:
\begin{itemize}[nosep,leftmargin=*]
\item A group of AI-human conversations sharing one failure mode
\item A list of EXISTING queries already in the benchmark
\end{itemize}

Tasks:
\begin{enumerate}[nosep,leftmargin=*]
\item Write one NEW retrieval query a researcher might use to find these conversations
\begin{itemize}[nosep,leftmargin=*]
\item Natural phrasing: ``Find conversations where the AI...''
\item Must capture the ABSTRACT failure pattern, not surface content
\item Must NOT use words from the descriptions verbatim
\item Must be discriminative: specific enough to exclude unrelated failures
\item Must NOT overlap with any existing query's failure mode
\end{itemize}
\item If you produced a query, grade each conversation's relevance (2 = central, 1 = tangential)
\end{enumerate}

If you cannot write a query that is clearly distinct from all existing queries,
set \texttt{query} to \texttt{null}. Dropping a near-duplicate is correct.

\textbf{User:} FAILURE TYPE: [canonical label]\\
CONVERSATIONS: [member conversations with descriptions]\\
EXISTING QUERIES: [current benchmark queries]
\end{promptbox}

\subsubsection{Stage 5: Pool and Expand}

\begin{promptbox}{Pooled Relevance Judgment (WildChat)}
\textbf{System:} You are an expert at identifying LLM failure modes in human-AI conversations.

You will receive a QUERY describing a very specific LLM failure pattern, and a set of
CANDIDATE CONVERSATIONS that have not yet been judged for relevance.

A candidate is relevant ONLY IF:
\begin{enumerate}[nosep,leftmargin=*]
\item The user's instruction matches the type of constraint described in the query
\item The AI's response violates that constraint in the specific way the query describes
\item A reasonable person would agree the failure is the same, not merely analogous
\end{enumerate}

Additional guidelines:
\begin{itemize}[nosep,leftmargin=*]
\item A candidate can be relevant even if the deviation appears unintentional or minor---what matters is whether the output differs from the exact specification
\item When instructions contain errors, judge against what the user actually specified
\end{itemize}

A candidate is NOT relevant if:
\begin{itemize}[nosep,leftmargin=*]
\item The AI makes a different kind of mistake
\item The failure mechanism differs from what the query describes
\item The AI actually complies correctly
\end{itemize}

Scoring: 2 = clearly relevant, 1 = marginally relevant, 0 = not relevant

Be conservative. When in doubt, assign 0.

\textbf{User:} QUERY: [query]\\
CONVERSATIONS TO JUDGE: [candidate conversations]
\end{promptbox}

\subsection{Math Meta-Program: Analogues via Shared Reasoning Technique}
\label{sec:appendix-math}

The Math Meta-Program benchmark targets problems sharing the same abstract proof strategy across completely different mathematical fields.

\subsubsection{Stage 2: Fingerprinting}

GPT-5 analyzes each problem and its solution to extract a reasoning fingerprint.

\begin{promptbox}{Reasoning Fingerprint Extraction}
\textbf{System:} You are a mathematics professor who has spent decades studying how
mathematical reasoning recurs across competitions, fields, and difficulty levels.

Your task: read a problem and its solution(s), then extract a REASONING FINGERPRINT
that captures the abstract cognitive move required---the ``aha moment''---stated so
domain-independently that a professor would recognize the same move in a problem
from a completely different field.

\textbf{CRITICAL: GRANULARITY OF meta\_strategy AND fingerprint\_summary}

These fields are the PRIMARY CLUSTERING KEYS. They must be abstract enough that
problems from completely different fields using the same reasoning move land in
the same cluster.

RULE: The fingerprint\_summary must be abstract enough to match a family of
problems, specific enough to exclude unrelated reasoning moves.

TOO COARSE (useless): ``induction'', ``pigeonhole'', ``contradiction''\\
TOO FINE (useless): anything that describes exactly one problem

\textbf{Output fields:}
\begin{itemize}[nosep,leftmargin=*]
\item \texttt{meta\_strategy}: The abstract reasoning move, the eureka. Use NO domain
      vocabulary. Must describe the logical maneuver so that a problem from a
      different field using the same move would fit.
\item \texttt{abstract\_proof\_move}: Domain-independent logical skeleton. Examples:
      ``sum over all rotations $\rightarrow$ global average $\rightarrow$ existence by integrality'' |
      ``assume extremal $\rightarrow$ local exchange argument $\rightarrow$ contradiction''
\item \texttt{key\_insight}: One sentence: the non-obvious observation that unlocks the problem.
      State WITHOUT domain vocabulary.
\item \texttt{fingerprint\_summary}: $\leq$20 words. The meta-program label for this family.
      PRIMARY CLUSTERING KEY.
\item \texttt{technique\_family}: algebra | combinatorics | geometry | number\_theory |
      real\_analysis | linear\_algebra | probability | game\_theory
\item \texttt{difficulty\_tier}: easy | medium | hard
\end{itemize}

\textbf{User:} PROBLEM: [problem statement]\\
SOLUTION: [solution text]
\end{promptbox}

\subsubsection{Stage 3: Clustering}

\begin{promptbox}{Meta-Program Label Normalization}
\textbf{System:} You are normalizing a list of meta-program labels from mathematical
reasoning analysis.

Many labels express the same underlying reasoning move with slightly different wording.
Collapse near-duplicates into a single canonical label. Keep genuinely distinct
reasoning moves as separate canonicals.

CRITICAL DISTINCTION: Two labels should collapse only if a mathematician would say
``yes, these require the same core aha moment''---NOT merely that they use the same
technique family, topic area, or proof technique.

Rules:
\begin{itemize}[nosep,leftmargin=*]
\item Every input label must appear exactly once as a key
\item Pick the most abstract, clearly-phrased version as canonical
\item Do NOT collapse labels that are merely in the same technique family
\end{itemize}

\textbf{User:} [List of meta-program labels to normalize]
\end{promptbox}

\begin{promptbox}{Cluster Merge Identification}
\textbf{System:} You are an expert mathematician reviewing clusters of math problems
grouped by their meta-program---the abstract reasoning move required to solve them.

You will receive a list of clusters, each with a canonical label and sample
fingerprint\_summaries of member problems.

Identify which clusters should be MERGED because they represent the same fundamental
reasoning pattern---the same abstract ``aha moment''---even if described differently.

DO NOT merge clusters that merely belong to the same technique family (e.g., ``all
use induction'') but require genuinely different insights.

Return merges as: \texttt{\{"merges": [\{"merge\_ids": [...], "new\_label": "...", "rationale": "..."\}]\}}

\textbf{User:} [List of clusters with labels and sample fingerprints]
\end{promptbox}

\subsubsection{Stage 3b: Cluster Validation}

\begin{promptbox}{Cluster Membership Verification}
\textbf{System:} You are an expert mathematician who specializes in recognizing when
problems require the same abstract reasoning pattern---the same ``aha moment''---even
across completely different fields.

You will receive a group of problems automatically clustered as sharing the same
meta-program. Your job is to verify which ones truly belong.

A problem belongs in the cluster if and only if:
\begin{itemize}[nosep,leftmargin=*]
\item Its core proof insight is structurally the same as the majority of other members
\item A student who mastered the pattern would immediately recognize the connection,
      even if topics look completely different
\end{itemize}

Problems in the same cluster CAN be from varying mathematical fields---it's the
reasoning style that matters.

A problem does NOT belong if:
\begin{itemize}[nosep,leftmargin=*]
\item It uses a similar technique family but the specific aha moment is different
\item Its fingerprint was accidentally grouped due to similar wording but the
      underlying reasoning move is distinct
\end{itemize}

Flag for removal only problems where you are confident the core insight is genuinely
different. When in doubt, keep.

Return: \texttt{\{"keep": [...], "remove\_candidates": [...], "canonical\_label": "...", "rationale": "..."\}}

\textbf{User:} CLUSTER LABEL: [label]\\
MEMBER PROBLEMS: [problems with fingerprints]
\end{promptbox}

\subsubsection{Cluster Selection}

\begin{promptbox}{Diversity-Based Cluster Selection}
\textbf{System:} You are helping curate a math reasoning-analogue retrieval benchmark.

You will receive descriptions of clusters of math problems. Each cluster groups
problems sharing the same abstract reasoning pattern regardless of field.

Task: Greedily select clusters that maximize diversity.

WITHIN-CLUSTER SURFACE DIVERSITY (makes individual queries hard):
A cluster is STRONG if its members span DIFFERENT mathematical fields (geometry,
number theory, analysis, etc.). This forces retrievers to understand the abstract
reasoning pattern, not match surface keywords.

Each cluster shows within-diversity = HIGH (3+ fields) / MED (2) / LOW (1).
HIGH and MED clusters should make up the bulk of your selection.

STOPPING RULE: Stop when all remaining candidates have LOW within-diversity.

\textbf{User:} [Cluster descriptors with diversity ratings]
\end{promptbox}

\subsubsection{Stage 5: Pool and Expand}

\begin{promptbox}{Pooled Relevance Judgment (Math)}
\textbf{System:} You are judging relevance for a mathematical reasoning-analogue
retrieval benchmark.

You will receive:
\begin{itemize}[nosep,leftmargin=*]
\item A QUERY PROBLEM and its reasoning fingerprint
\item GOLD PROBLEMS confirmed to share its meta-program (for calibration)
\item CANDIDATE PROBLEMS that are unjudged
\end{itemize}

For each candidate, decide if it shares the same CORE REASONING PATTERN---the same
abstract ``aha moment''---regardless of topic, vocabulary, or mathematical field.

Relevance grades:
\begin{itemize}[nosep,leftmargin=*]
\item 2 = SAME meta-program: a mathematician would immediately recognize the same
      eureka insight across different fields. Abstract proof move is structurally identical.
\item 1 = ADJACENT reasoning: genuinely related proof strategy but a distinct core
      insight. A student who mastered the query would be helped but not immediately equipped.
\item 0 = DIFFERENT reasoning pattern entirely.
\end{itemize}

\textbf{User:} QUERY: [problem with fingerprint]\\
GOLD: [confirmed analogues]\\
CANDIDATES: [problems to judge]
\end{promptbox}

\subsubsection{Evaluation: Listwise Ranking}

\begin{promptbox}{Oracle GPT 5.2 Tournament Ranking with no solutions}
\textbf{System:} You are an expert mathematician. Your task is to identify problems
that share the same ABSTRACT PROOF STRATEGY, i.e., the same ``meta-move'', even when topics look completely different.

IMPORTANT: You must INFER how each problem would be solved. You don't have access
to solutions---you must figure out the likely proof approach from the statement alone.

WHAT ``SAME ABSTRACT STRATEGY'' MEANS:

Two problems match if their solutions would use the same high-level meta-move:
\begin{itemize}[nosep,leftmargin=*]
\item ``Exploit uniqueness of X to constrain the answer''
\item ``Transform representation $\rightarrow$ swap order of operations $\rightarrow$ collapse''
\item ``Diagonalize/decompose $\rightarrow$ solve per component $\rightarrow$ reassemble''
\item ``Finite domain: injective $\Leftrightarrow$ surjective''
\item ``Symmetry averaging to project onto invariants''
\end{itemize}

Problems can share a meta-move despite having NOTHING in common on the surface.

IGNORE surface similarity: same field, same objects, similar notation, problems that ``look alike''

FOCUS ON deep proof skeleton: What KEY INSIGHT unlocks the problem? What meta-level
principle does the solution rely on? Would the same proof OUTLINE work for both?

Rank ALL candidates. Return: \texttt{\{"ranked": [\{"candidate\_num": N, "reason": "..."\}, ...]\}}

\textbf{User:} QUERY: [problem statement]\\
CANDIDATES: [numbered list of problems]
\end{promptbox}

\begin{promptbox}{Oracle GPT 5.2 Tournament Ranking with solutions}
\textbf{System:} You are an expert mathematician. Your task is to identify problems
that share the same ABSTRACT PROOF STRATEGY, i.e., the same ``meta-move''.

You are given BOTH problem statements AND solutions. Use the solutions to identify
the core reasoning technique, not just surface similarity.

IMPORTANT: Mathematical problems often have MULTIPLE valid solution approaches. Two
problems may share a reasoning pattern via one proof method even if other solutions
exist. Consider ALL plausible strategies:
\begin{itemize}[nosep,leftmargin=*]
\item A counting problem might use bijection, generating functions, or recursion
\item A number theory problem might use modular arithmetic, induction, or construction
\item An inequality might yield to AM-GM, Cauchy-Schwarz, or calculus
\end{itemize}

If ANY natural solution approach shares the same core insight, treat them as matching.

FOCUS ON the deep proof skeleton revealed in solutions: What KEY INSIGHT unlocks each?
What meta-level principle does each rely on? Do solutions follow the same structural pattern?

\textbf{User:} QUERY: [problem with solution excerpt]\\
CANDIDATES: [problems with solution excerpts]
\end{promptbox}

\begin{promptbox}{Two-Stage Think-First Ranking}
\textbf{System:} You are an expert mathematician identifying problems with shared
abstract proof strategy.

IMPORTANT: Work in TWO STAGES:

\textbf{STAGE 1 - THINK FIRST:}
Before ranking, reason through how each problem would be solved. For the query and
each candidate: What's the key insight? What proof technique? What's the abstract meta-move?

\textbf{STAGE 2 - RANK:}
Only after thinking through each problem's approach should you rank by whether they
share the query's abstract proof strategy.

Return:
\begin{verbatim}
{
  "query_analysis": {"key_insight": "...", "meta_move": "..."},
  "candidate_analyses": [{"candidate_num": N, "key_insight": 
  
  "...", "meta_move": "..."}, ...],
  "ranked": [{"candidate_num": N, "reason": "..."}, ...]
}
\end{verbatim}

\textbf{User:} QUERY: [problem]\\
CANDIDATES: [numbered problems]
\end{promptbox}

\subsection{Writing-Style: Analogues via Cross-Domain Authorship}
\label{sec:appendix-writing}

The Writing-Style benchmark targets authorship identification across unrelated topics,
testing whether stylistic invariants persist when topic is factored out. Unlike other
benchmarks, Writing-Style skips Stages 2--3 (authorship is ground truth) and proceeds
directly to evaluation.

\subsubsection{Evaluation: Listwise Ranking}

\begin{promptbox}{Oracle GPT 5.2 Tournament Ranking (Authorship)}
\textbf{System:} You are an expert in authorship analysis and stylometry.

You will receive a QUERY SNIPPET and CANDIDATE SNIPPETS. Rank candidates by how
likely they were written by the same author as the query.

Focus on writing style, NOT topic:
\begin{itemize}[nosep,leftmargin=*]
\item Sentence rhythm and syntactic patterns
\item Hedging language and epistemic stance
\item Characteristic vocabulary and phrasing
\item Tone and rhetorical posture
\end{itemize}

The same author may write about different topics---topic similarity is NOT the
signal you should rely on.

Every candidate number MUST appear exactly once.

Return: \texttt{\{"ranked": [\{"candidate\_num": N, "reason": "<brief style observation>"\}, ...]\}}

\textbf{User:} QUERY SNIPPET:\\
"""[query text]"""

CANDIDATE SNIPPETS: [numbered snippets]

Rank by stylistic similarity to the query author.
\end{promptbox}

\subsubsection{Evaluation: Multi-Hop Retrieval}

\begin{promptbox}{Multi-Hop Query Generation (Authorship)}
\textbf{System:} You are an expert at author attribution, i.e, identifying texts written by the same author.

Given an original text snippet and notes from previous search iterations, generate
a search query that will help find other texts written by the same author.

The search query should:
\begin{enumerate}[nosep,leftmargin=*]
\item Focus on distinctive stylistic features, vocabulary patterns, or thematic preferences
\item Capture the author's unique voice and writing mannerisms
\item Be different from previous search angles to maximize coverage
\end{enumerate}

If this is the first hop, focus on the most distinctive stylistic markers. For
later hops, refine based on patterns you've discovered.

\textbf{User:} ORIGINAL TEXT: [query snippet]\\
NOTES FROM PREVIOUS HOPS: [accumulated observations]\\
HOP NUMBER: [N] of [total]
\end{promptbox}

\begin{promptbox}{Multi-Hop Note Extraction (Authorship)}
\textbf{System:} You are an expert at author attribution.

Your tasks:
\begin{enumerate}[nosep,leftmargin=*]
\item Select text snippets that appear to be written by the SAME AUTHOR as the query
\item Write brief notes about the stylistic patterns you observed
\end{enumerate}

Look for: vocabulary choices, sentence structure, punctuation habits, thematic
preferences, tone, rhetorical devices, and other authorial fingerprints.

\textbf{User:} QUERY TEXT: [snippet]\\
PREVIOUS NOTES: [observations]\\
CANDIDATES: [retrieved snippets]

Return: \texttt{\{"candidate\_ids": [...], "notes": "...", "summary": "..."\}}
\end{promptbox}

\begin{promptbox}{Query Rewriting (Authorship)}
\textbf{System:} You are an expert at author attribution.

Given a text snippet, rewrite the query to better capture the distinctive stylistic
features that would help find other texts by the same author. Focus on vocabulary
patterns, sentence structure, tone, and other authorial fingerprints.

Return: \texttt{\{"rewritten\_query": "...", "rationale": "..."\}}

\textbf{User:} ORIGINAL TEXT: [snippet]

Rewrite to capture the author's distinctive style.
\end{promptbox}

\subsection{Congress Hearings: Tip-of-the-Tongue Retrieval}
\label{sec:appendix-congress}

The Congress Hearings benchmark targets tip-of-the-tongue retrieval: matching fuzzy,
impressionistic recollections to specific hearing passages. Unlike other benchmarks,
Congress skips Stage 3 (each query targets a single passage rather than a cluster).

\subsubsection{Stage 2: Memorability Rating and Query Generation}

\begin{promptbox}{ToT Query Generation from Memorable Passages}
\textbf{System:} You are helping construct a ``Tip of the Tongue'' (ToT) information
retrieval benchmark over US congressional hearing transcripts.

You will be shown a passage from a hearing---typically a Q\&A exchange between a
legislator and a tech industry witness.

\textbf{Your job:}

\textbf{1. RATE the passage's memorability (1-5):}
\begin{itemize}[nosep,leftmargin=*]
\item 1 = boring procedural, nobody would remember
\item 2 = mildly interesting but generic
\item 3 = somewhat memorable, has a specific detail worth recalling
\item 4 = very memorable, a distinct confrontation or revelation
\item 5 = iconic, widely reported moment
\end{itemize}

\textbf{2. If memorability $\geq$ 3, write a ToT POST} ($\sim$200 words, written as
someone posting on Reddit trying to recall this moment):

\textbf{MUST FOLLOW:}
\begin{itemize}[nosep,leftmargin=*]
\item Do NOT include names of any person, company, platform, committee, or legislation
\item Do NOT include dates, years, or exact identifiers
\item Reflect imperfect memory: mix up minor details, be uncertain about specifics,
      conflate with adjacent moments. Do NOT explicitly say ``I'm not sure''---state
      distorted details naturally
\item Describe what happened in indirect, experiential terms---what it felt like,
      the dynamic, body language, tone
\item Natural conversational tone. No formal writing. Write like you're posting on Reddit
\item Skip greetings like ``Hey everyone''---begin directly with your memory
\item End with an open-ended question inviting others to help find it
\item The post must be genuinely hard for search to match. Avoid verbatim words from the passage
\end{itemize}

\textbf{COULD FOLLOW:}
\begin{itemize}[nosep,leftmargin=*]
\item Set the scene with a personal anecdote about when/where you encountered this
\item Focus on emotional impact---tension, awkward silence, audible reaction
\item Make subtle comparisons to similar moments without naming them
\item Include 1-2 plausible but INCORRECT details (memory distortion)
\item Mention sensory details: camera zoom, sounds from gallery, fidgeting
\item Reference the emotional arc: buildup, confrontation, deflection
\end{itemize}

\textbf{BAD} (keywords overlap, too specific):\\
``When did the Facebook CEO testify about data privacy before the Senate?''

\textbf{GOOD} ($\sim$200 words, oblique, personal, distorted):\\
``So I was procrastinating at work a few years back and ended up watching one of
those government hearings where they bring in tech people, and there was this one
exchange that's stuck with me. One of the older guys on the panel---I want to say
he was from somewhere in the south but I might be mixing him up---he pulled out
what looked like a printout of something from inside the company itself. Like
internal stuff. And he read part of it out loud, and the exec at the table just
had this deer-in-headlights moment...''

\textbf{User:} PASSAGE: [hearing transcript excerpt]
\end{promptbox}

\subsubsection{Opening Style Diversification}

To prevent mode collapse in query formulation, we diversify opening styles using
15 distinct patterns:

\begin{promptbox}{Opening Diversification}
\textbf{System:} You are rewriting forum posts to have more diverse openings.

You will receive a $\sim$200 word forum post where someone is trying to recall a
congressional hearing moment. The content and meaning must stay EXACTLY the same.
Your ONLY job is to rewrite the opening 1-2 sentences to use this specific style:

STYLE: [one of the styles below]

Rules:
\begin{itemize}[nosep,leftmargin=*]
\item Keep the EXACT same content, details, memory distortions, and closing question
\item ONLY change the first 1-2 sentences to match the requested style
\item Do NOT add or remove any information
\item Output ONLY the rewritten post
\end{itemize}

\textbf{Opening styles used:}
\begin{enumerate}[nosep,leftmargin=*]
\item Frustrated question (``This has been driving me crazy...'')
\item Mid-thought, no preamble (``Ok so there was this hearing where...'')
\item Setting a scene (``I was at my desk / on the couch...'')
\item A comparison (``It was kind of like that other time when...'')
\item Challenge to the reader (``Does anyone else remember...'')
\item Stating what stuck (``The thing that always stuck with me was...'')
\item Diving straight in (``There's this clip where...'')
\item Temporal anchoring (``A couple years back, maybe around election season...'')
\item Explaining why you're posting (``My coworker mentioned something today...'')
\item Emotional reaction first (``I still get secondhand embarrassment...'')
\item A disclaimer (``I might be mixing up two different things here but...'')
\item Referring to how you saw it (``Someone sent me a clip once of...'')
\item Strong opinion opener (``Honestly one of the wildest moments...'')
\item A question to yourself (``Why can I never find this clip again?'')
\item Anchoring to another memory (``So right around the time that scandal...'')
\end{enumerate}
\end{promptbox}

\subsubsection{Evaluation: Query Rewriting}

\begin{promptbox}{ToT Query Rewriting for Transcript Matching}
\textbf{System:} You are an advanced retrieval system. You will be given a tip of tongue query
describing a user's hazy memory of a specific moment from a US congressional hearing.
They wrote a vague description of what they remember.

Rewrite their description as a search query that is more likely to match the
actual hearing transcript. Use the kind of language that would appear in an
official transcript based on whatever you can infer from their description.

Return ONLY a JSON object:
\begin{verbatim}
{
  "rewritten_query": "<your rewritten query>",
  "reasoning": "<what you think they're describing and why>"
}
\end{verbatim}

\textbf{User:} \{query\}

Rewrite this to better match the actual hearing transcript.
\end{promptbox}

\subsubsection{Evaluation: Multi-Hop Retrieval}

\begin{promptbox}{Multi-Hop Query Generation (Congress ToT)}
\textbf{System:} You are an advanced retrieval system. You will be given a tip of tongue query
describing a user's hazy memory of a specific moment from a US congressional hearing.
They wrote a vague description of what they remember. Given their
description and any notes from previous search attempts, generate a search query
that would match the actual hearing transcript.

Think about what the transcript would actually contain---speaker names, committee
procedures, policy terms, specific phrases.

\textbf{User:} ORIGINAL DESCRIPTION:\\
\{original\_query\}

NOTES FROM PREVIOUS HOPS:\\
\{notes\}

HOP: \{hop\_num\} of \{total\_hops\}

Generate a search query. Return ONLY JSON:
\begin{verbatim}
{
  "search_query": "<query using transcript-language>",
  "rationale": "<what you're looking for this hop>"
}
\end{verbatim}
\end{promptbox}

\begin{promptbox}{Multi-Hop Note Extraction (Congress ToT)}
\textbf{System:} You are trying to find a specific congressional hearing moment that someone vaguely
remembers. You will see their description and some candidate transcript passages.

Select any passages that could plausibly be the moment they're describing.
Write notes about what you've learned so far---what matches, what doesn't,
what to search for next.

\textbf{User:} WHAT THEY REMEMBER:\\
\{query\}

PREVIOUS NOTES:\\
\{previous\_notes\}

CANDIDATE PASSAGES (hop \{hop\_num\}):\\
\{candidates\}

Analyze these passages. Return ONLY JSON:
\begin{verbatim}
{
  "candidate_ids": ["id1", "id2", ...],
  "notes": "<what matched, what didn't, what to try next>",
  "summary": "<brief summary of this hop>"
}
\end{verbatim}

\{selection\_instruction\}
\end{promptbox}

\begin{promptbox}{Multi-Hop Reranking (Congress ToT)}
\textbf{System:} You are trying to find a specific congressional hearing moment. Rank passages
by how likely they are the moment being described. Output exactly \{k\} items. Every candidate number must appear exactly once. Return ONLY JSON:
\begin{verbatim}
{
  "ranked": [
    {"candidate_num": <int>, "reason": "<brief>"},
    ...
  ]
}
\end{verbatim}

\textbf{User:} WHAT THEY REMEMBER:\\
\{query\}

CANDIDATE PASSAGES:\\
\{candidates\}

Rank all \{k\} candidates. Every number must appear exactly once.
\end{promptbox}

\section{Offline-Indexing}
\label{offline-indexing}
We test how much of the latent property a query-agnostic index can capture offline. On Math, Opus~5.5 writes a 100--200 word description of each problem from a fixed prompt that asks
for its topic, likely techniques, and structure; the model never sees queries, relevance labels,
or construction attributes (see the prompt in Figure~\ref{fig:offline-prompt}). We append each description to its problem, and, since Math queries
are themselves problems, append the description of each query's own problem to the query. As an
oracle, we do the same with the fingerprints produced during benchmark construction, which systems
never receive. All rows use Qwen3-Embedding-4B with the setup of Table~\ref{tab:analogue-results}.

Table~\ref{tab:latent-index} shows the results. The offline index recovers 23--35\% of the
oracle's gain over the baseline. Because the fingerprints seeded the clusters behind the relevance
labels, the oracle row is an upper bound rather than an achievable target.

\begin{figure}[t]
\begin{tcolorbox}[colback=gray!5, colframe=gray!60, boxrule=0.5pt, arc=2pt,
                  left=6pt, right=6pt, top=4pt, bottom=4pt,
                  title=\small Offline describer prompt (Claude Opus 5.5)]
\small\ttfamily
\textbf{System:} You write entries for a search index over a collection of math problems. You
will not see the searches that will be run against the index, so describe each problem in a way
that would help a wide range of possible searches find it.

\medskip
\textbf{User:} Here is a math problem:

\smallskip
\textless problem\textgreater\\
\{problem\}\\
\textless /problem\textgreater

\smallskip
Write a search-index description of this problem in 100 to 200 words of plain prose. Cover its
area of mathematics and topic, the concepts, theorems, and techniques a solution would likely use,
what the problem asks for and how it is structured, and any other properties someone searching a
collection of math problems might look for. Do not include a full solution. Output only the
description.
\end{tcolorbox}
\caption{Prompt for the offline index (Appendix~\ref{offline-indexing}). It is run once per
problem, with \texttt{\{problem\}} replaced by the problem statement as it
appears in the corpus.}
\label{fig:offline-prompt}
\end{figure}

\begin{table}[t]
\caption{Offline indexing on Math (gold/pooled), using Qwen3-Embedding-4B and the setup of
Table~\ref{tab:analogue-results}. Each problem and each query is expanded either with an LLM
description written offline, without seeing queries or labels, or with its fingerprint from
benchmark construction. The fingerprint row is an oracle, since fingerprints seeded the relevance labels. Descriptions recover 23--35\% of the fingerprints' gain.}
\centering
\small
\setlength{\tabcolsep}{4pt}
\begin{tabular}{l cc cc ccc}
\toprule
& \multicolumn{2}{c}{NDCG@10} & \multicolumn{2}{c}{NDCG@50} & R@10 & R@50 & R@100 \\
\cmidrule(lr){2-3} \cmidrule(lr){4-5}
Pipeline & G & P & G & P & G/P & G/P & G/P \\
\midrule
Qwen3-Embed-4B & \hgold{.098} & \hpooled{.130} & \gold{.132} & \pooled{.162} & \cellcolor{highlight!30}.089/.110 & .200/.224 & .287/.318 \\
\quad + offline descriptions & \hgold{.138} & \hpooled{.177} & \gold{.193} & \pooled{.234} & \cellcolor{highlight!30}.124/.151 & .298/.337 & .410/.440 \\
\cmidrule{1-8}
\quad + construction fingerprints & \hgold{.276} & \hpooled{.320} & \gold{.347} & \pooled{.387} & \cellcolor{highlight!30}.238/.263 & .525/.547 & .653/.668 \\
\bottomrule
\end{tabular}
\label{tab:latent-index}
\end{table}

\section{Statistical Uncertainty of the Retrieval--Verification Gap}
\label{app:bootstrap}

To check that the retrieval--verification gap is not driven by a few queries, we compute
paired bootstrap intervals over queries. For each task, we take the per-query difference in NDCG@10 between the oracle and a compared system, resample the queries with replacement 10,000 times, and report the 2.5th and 97.5th percentiles of the mean difference. Both systems are always scored on the same resampled queries, and every comparison uses the same resamples. Table~\ref{tab:bootstrap} shows that every interval excludes zero, against both the best
non-oracle system and Gemini-Embedding-2, and that the oracle scores higher on most
individual queries in every task. The intervals are widest on WildChat Errors, which has
40 queries.
\begin{table}[ht]
\centering
\caption{\textbf{The retrieval--verification gap is consistent across queries.}
Each gap is the oracle's NDCG@10 minus the compared system's, averaged over queries, with
a 95\% interval from a paired bootstrap: the queries are resampled with replacement 10,000
times, and both systems are scored on the same resamples. H\,/\,L\,/\,T counts the queries
on which the oracle's NDCG@10 is higher than, lower than, or tied with the system's. Each
task compares the oracle with its best non-oracle system and with Gemini-Embedding-2, the
strongest single-stage retriever; on Writing-Style these are the same system. Writing-Style
and Congress Hearings have no pooled labels. The Math oracle is the GPT-5.2 Tournament
without solutions.}
\label{tab:bootstrap}
\small
\setlength{\tabcolsep}{4pt}
\begin{tabular}{@{}llcccc@{}}
\toprule
 & & \multicolumn{2}{c}{Gold} & \multicolumn{2}{c}{Pooled} \\
\cmidrule(lr){3-4} \cmidrule(l){5-6}
Task {\footnotesize(queries)} & Compared with & Gap [95\% CI] & H\,/\,L\,/\,T & Gap [95\% CI] & H\,/\,L\,/\,T \\
\midrule
Twitter {\footnotesize(281)} & GPT-5.2 Multi-Hop Agent & .190 [.159, .220] & 204\,/\,45\,/\,32 & .253 [.224, .281] & 242\,/\,32\,/\,7 \\
 & Gemini-Embedding-2 & .270 [.239, .302] & 231\,/\,19\,/\,31 & .343 [.312, .372] & 256\,/\,21\,/\,4 \\
\midrule
WildChat {\footnotesize(40)} & GPT-5.2 Multi-Hop Agent & .332 [.231, .436] & 28\,/\,3\,/\,9 & .327 [.228, .426] & 29\,/\,7\,/\,4 \\
 & Gemini-Embedding-2 & .342 [.234, .447] & 29\,/\,2\,/\,9 & .336 [.235, .435] & 32\,/\,4\,/\,4 \\
\midrule
Math {\footnotesize(151)} & GPT-5.2 Multi-Hop Agent & .114 [.085, .145] & 91\,/\,22\,/\,38 & .118 [.086, .151] & 89\,/\,28\,/\,34 \\
 & Gemini-Embedding-2 & .124 [.088, .160] & 95\,/\,25\,/\,31 & .121 [.084, .159] & 92\,/\,31\,/\,28 \\
\midrule
Writing {\footnotesize(512)} & Gemini-Embedding-2 & .351 [.325, .377] & 414\,/\,41\,/\,57 & -- & -- \\
\midrule
Congress {\footnotesize(254)} & GPT-5.2 Multi-Hop Agent & .730 [.677, .780] & 197\,/\,2\,/\,55 & -- & -- \\
 & Gemini-Embedding-2 & .854 [.814, .892] & 232\,/\,0\,/\,22 & -- & -- \\
\bottomrule
\end{tabular}
\end{table}

\section{Implementation \& Hardware Details}
\label{sec:appendix-impl}
\subsection{Embedding and lexical retrieval} Dense-embedding inference for Qwen3-Embedding-0.6B and Qwen3-Embedding-4B was performed on a single H200 GPU; LateOn (149M parameters) was indexed and queried with PyLate using PLAID, also on a single H200. BM25 indexing and retrieval ran on CPU using the rank-bm25 Python library. None of these workloads are demanding for a retrieval benchmark of this size — the largest corpus is the 507K-conversation WildChat split, and the others are smaller — so any modern data-center GPU (e.g., A100, H100, L40) is sufficient to reproduce the embedding-side results, and the CPU lexical baseline can be reproduced on a commodity laptop.

\subsection{Commercial APIs} Benchmark construction (annotation, label collapse, query generation, and pooled-judgment expansion), the GPT-5.2 Query Rewriter, the four-hop GPT-5.2 Multi-Hop Agent, and the Oracle GPT-5.2 Tournament all rely on commercial APIs: OpenAI for GPT-5, GPT-5.2, and GPT-5.4-nano, and Google for Gemini-Embedding-2. Construction-stage sweeps used GPT-5.4-nano where possible (notably the WildChat corpus-wide failure-mode pass) for cost reasons, while query writing, cluster verification, and tournament reranking used GPT-5 / GPT-5.2. All API calls were made with semaphore-capped concurrency to stay within rate limits.

\subsection{Reproducibility} All retriever hyperparameters appear in section 5; the tournament reranking algorithm (batch size b=20, promotion k=4) is described in Appendix~\ref{sec:tournament}; and the construction-stage prompts that drive the API calls are reproduced verbatim in Appendix~\ref{sec:appendix-prompts}. The HuggingFace repository contains the
exact corpora, queries, and relevance judgments evaluated in this paper, and the GitHub repository contains the construction and evaluation code, the prompts, and the rankings produced by every system in Tables~2--4. The embedding and BM25 numbers are therefore reproducible, and every reported
number can be rescored from the released rankings. The LLM-based pipelines are stochastic, but cell-level metrics aggregate over hundreds of query--document judgments per task (see \S7 of the checklist).

\section{Limitations}
\label{sec:app-limits}
\subsection{Reliance on frontier-LLM judges}
Both the construction pipeline (Stages 2, 3, 4, and 5 in section 4) and the verification ceiling in our retrieval–verification diagnostic (see section 3) rely on frontier reasoning LLMs (GPT-5, and Claude Opus 5.5 for the label audit). If these models share systematic blind spots in how they read implicit stance, behavioral failures, proof strategies, or stylistic cues, those biases propagate into gold labels and into the upper bound ($V_t$). The retrieval–verification gap is therefore best read as a gap relative to a specific verifier class, not as a model-independent claim about retrievability.

\subsection{Scale and statistical power of individual tasks} Several tasks have modest query counts (Table 1: WildChat with 40 queries, Math Meta-Program with 151, Twitter-Conflict with 281). This is a deliberate consequence of high-recall judging over latent attributes, but it limits the resolution at which fine-grained comparisons between similarly performing retrievers can be drawn. 

\subsection{Coverage and domain breadth} Each oblique mechanism is instantiated by one or two corpora. The Twitter-Conflict corpus is scoped to a single geopolitical conflict beginning March 2026, the Writing-Style corpus draws on 64 researchers, and the Congress Hearings corpus is U.S.-specific and English-only. We do not evaluate cross-language generalization or generalization across implicit-stance topics, authorship populations, or legislative bodies. Our claims about obliqueness should therefore be read as evidence for the existence and difficulty of these regimes, rather than as comprehensive coverage of them.

\subsection{Cost of the verification ceiling} The Oracle GPT-5.2 Tournament is not a scalable retriever — it is an upper-bound estimator that performs many listwise LLM calls per query. As a result, the bound it establishes is currently dependent on access to frontier APIs, which limits both reproducibility under tight compute budgets and the practicality of using the tournament as a reranker in deployment. Improving the quality/cost tradeoff of this ceiling is itself an open problem, as noted in section 5.2.

\subsection{Construction-pipeline assumptions} Our pipeline assumes that a human-defined lens (f) plus an LLM annotator can extract a meaningful latent attribute in a single corpus pass, and that clustering attribute values surfaces queries that real users would also pose. Both assumptions are imperfect: the lens may bias which kinds of obliqueness are surfaced, and clusters that emerge from LLM annotations may not align with the queries a downstream practitioner would actually write. Writing-Style and Congress Hearings sidestep this by using ground-truth labels (authorship and known target passages, respectively), but the descriptive and analogue tasks remain pipeline-dependent.

\section{Broader Impacts}
\label{sec:app-broader}
OBLIQ-Bench is a retrieval evaluation benchmark, not a deployed system, and we expect its primary impact to be on retrieval-research methodology. The benchmark is designed to motivate scalable retrieval architectures that surface latent and implicit relevance, with downstream applications including auditing LLM behavior at scale (WildChat Errors), supporting accountability and journalistic search over public discourse (Congress Hearings, Twitter-Conflict), enabling training-data decontamination via shared-reasoning analogue retrieval (Math Meta-Program), and exploring stylometric search across unrelated topics (Writing-Style).

We also note dual-use concerns. Improvements on implicit-stance retrieval could be repurposed for political surveillance over social-media corpora; better cross-domain stylometric retrieval could be used to deanonymize pseudonymous authors who write under multiple personas; and behavioral-failure-mode retrievers built on top of WildChat Errors could be used adversarially to mine examples of model misbehavior. We mitigate these risks at the dataset level rather than at the model level, and we release data, judgments, code, prompts, and system rankings, but no trained classifiers or fine-tuned retrievers, which keeps the marginal contribution toward malicious uses small. A more detailed Responsible AI statement is included in the Croissant metadata file accompanying the release.

\section{Licenses and Provenance}
\label{sec:app-license}
All assets used in OBLIQ-Bench are credited and used within their stated terms.

Corpora. The WildChat-4.8M corpus  is used under its original release license. X/Twitter content was collected through the X API full-archive search endpoint under the developer terms of service for academic research; we release tweet text with author handles stripped, while @-mentions inside tweet bodies are preserved. Congressional hearing transcripts from the 110th–119th Congresses, supplemented by ten high-profile technology hearings, are sourced from GovInfo, which is U.S. public-domain material. Mathematics problems are drawn from publicly available archives of the Putnam Competition, related undergraduate competitions, the American Mathematical Monthly, and qualifying-exam sources. Writing-Style author snippets are derived from publicly posted blog content of 64 researchers with author identifiers scrubbed; length-matched distractors are sampled from LWN.net, LessWrong, and Quanta Magazine in compliance with each site's redistribution terms.

Models and software. BM25 is implemented via the rank-bm25 library. LateOn (149M parameters) is used via PyLate and the PLAID indexing algorithm. Qwen3-Embedding-0.6B and Qwen3-Embedding-4B are accessed under their published terms. Gemini-Embedding-2  is accessed via the Google API. GPT-5, GPT-5.2, and GPT-5.4-nano are accessed via the OpenAI API.Per-split license, version, and provenance information is documented in the Croissant metadata file in the HuggingFace repository.


\end{document}